\documentclass[12pt,letterpaper]{report}
\usepackage[activeacute,english]{babel}
\usepackage{amsmath}
\usepackage{amsbsy}
\usepackage{amsthm}
\usepackage{amssymb}
\usepackage{latexsym}

\begin{document}

\title{Instituto Superior de Ciencias y Tecnolog\'{\i}a Nucleares \\
\vspace{2cm} Dissertation Diploma Thesis \\ \vspace{2.5cm} {\bf
About an Alternative Vacuum State for \\ Perturbative QCD \\
 \vspace{1.7cm}}}

\author{{\bf Author: Marcos Rigol Madrazo} \\ \vspace{0.1cm} \and 
{\bf Advisor: Dr. Alejandro Cabo Montes de Oca}}

\date{\vspace{1cm} Havana 1999}

\maketitle

\begin{abstract}

A particular initial state for the construction of a perturbative
QCD expansion is investigated. It is formed as a coherent
superposition of zero momentum gluon pairs and shows Lorentz as
well as global $SU(3)$ symmetries. The general form of the Wick
theorem is discussed, and it follows that the gluon and ghost
propagators determined by the proposed vacuum state, coincides
with the ones used in an alternative of the usual perturbation
theory proposed in a previous work, and reviewed here. Therefore,
the ability of such a procedure of producing a finite gluon
condensation parameter already in the first orders of perturbation
theory is naturally explained. It also follows that this state
satisfies the physicality condition of the BRST procedure in its
Kugo and Ojima formulation. A brief review of the canonical
quantization for gauge fields, developed by Kugo and Ojima, is
done and the value of the gauge parameter $\alpha$ is fixed to
$\alpha=1$ where the procedure is greatly simplified. Therefore,
after assuming that the adiabatic connection of the interaction
does not take out the state from the interacting physical space,
the predictions of the perturbation expansion for the physical
quantities, at the value $\alpha=1$, should have meaning. The
validity of this conclusion solves the gauge dependence
indeterminacy remained in the proposed perturbation expansion.

\end{abstract}

\tableofcontents

\chapter{Introduction}

Quantum Chromodynamics (QCD) was discovered in the seventies and
it has been considered as the fundamental theory for the strong
interactions. A theory of such sort, showing a non abelian
invariance group, was first suggested by Yang and Mills
\cite{Yang}. The main idea in it is the principle of local gauge
invariance, which for example, in Quantum Electrodynamics (QED)
means that the phase of the wave function can be defined in an
arbitrary way at any point of the space-time. In a non abelian
theory, the arbitrary phase is generalized to an arbitrary
transformation in an internal symmetry group, for QCD the internal
symmetry group is $SU(3)$. The discovery of this theory generated
radical changes in the character of the Modern Theoretical Physics
and as a consequence it has been deeply investigated in the last
years. It is believed at the present time that all the
interactions in Nature are gauge invariant \cite{Green}.

In one limit, the smallness of the coupling constant at high
momentum (asymptotic freedom) made possible the theoretical
investigation of the so-called hard processes by using the
familiar perturbative language. This so called perturbative QCD
(PQCD) was satisfactorily developed. However, relevant phenomena
associated with the strong interactions can't be described by the
standard perturbative methods and the development of the so-called
non-perturbative QCD is at the moment one of the great challenges
of Theoretical Physics.

One of the most peculiar characteristics of the strong
interactions is the color confinement. According to this
philosophy colored objects, like quarks and gluons, can't be
observed as free particles in contrast with hadrons that are
colorless composite states and effectively detected. The physical
nature of such phenomenon remains unclear. Qualitatively, it is
compared with the Meissner effect in superconductors, in which the
magnetic field is expelled from the bulk in which the Cooper pairs
condensate exists. It is considered that the QCD vacuum expels the
color fields from it. Numerous attempts to explain this property
have been made, for example explicit calculations in which the
theory is regularized in a spatial lattice \cite{Creutz}, and also
through the construction of phenomenological models. One of this
models, the so called MIT Bag Model \cite{Chodos}, assumes that a
bag or bubble is formed around the objects having color in such a
manner that they could not escape from it, because their effective
mass is smaller inside the bag volume and very high outside. The
dimensional quantity introduced in this model is $B$, the bag
constant, which is the pressure that the vacuum makes on the color
field. Another approach is the so called String Model
\cite{Gervais} which is based in the assumption that the
interaction forces between quarks and antiquarks grow with the
distance, in such a way that the energy increases linearly with
the string length $E(L)=kL$. The main parameter introduced in this
theory is the string tension $k$ which determines the strength of
the confining interaction potential.

A fundamental problem in QCD is the nature of its ground state
\cite{Shuryak}. This state is imagined as a very dense state of
matter, composed of gluons and quarks interacting in a complicated
way. Its properties are not easy accessible in experiments,
because quarks and gluons fields can't be directly observed, only
the color neutral hadrons are detected. Furthermore, the
interactions between quarks can't be directly determine, because
their scattering amplitudes can't be measured. It is known, from
the experience in solid-state physics, that a good understanding
of the ground state structure implies a natural explanation of
many of the phenomenological facts concerning to its excitations.
The theory of superconductivity is a good example, up to the
moment in which a good theory of the ground state was at hand the
description of its excitations remained basically
phenomenological.

It is already accepted that in QCD the zero-point oscillations of
the coupled modes produce a finite energy density, such effects
are called non-perturbative ones. Obviously such an energy density
can be subtracted by definition, however this procedure does not
solve the problem, because soft modes are rearranged in the
excited states and the variation of their energy should be
unavoidable considered. This energy density is determined
phenomenologically and its numerical estimate is \cite{Shuryak}
\[
E_{vac}\simeq -f\langle 0\mid g^2G^2\mid 0\rangle \simeq
0.5GeV/fm^3,
\]
where the so-called non-perturbative gluonic condensate $\langle
0\mid g^2G^2\mid 0\rangle$ was introduced and phenomenologically
evaluated by Shiftman, Vainshtein and Zakharov \cite{Zakharov}.
The negative sign of $ E_{vac}$ means that the non-perturbative
vacuum energy is lower that the one associated to the perturbative
vacuum.

Some of the QCD vacuum models, developed to explain the
above-mentioned properties, are mentioned below. These models can
be classified considering the dimension of the manifold in which
the non-perturbative field fluctuations are concentrated.

1- The ``instanton'' model, in which it is assumed that the field
is gathered in some localized regions of the space and time as
instantaneous fluctuations. These are considered as fluctuations
concentrated in zero dimensional manifolds.

2- The ``soliton'' model, in which it is assumed that the non
linear gauge fields create some kind of stable particles or
solitons (i.e. glueballs \cite{Hansson} or monopoles
\cite{Mandelstam}) in the space. The space-time manifold to be
considered for these models is one-dimensional.

3- The ``string'' model, in which closed strings (field created
between color charges shows a form resembling a flux tube or
string) are present in the vacuum. In space-time the history of
these strings is a 2-dimensional surface, so in this picture the
fluctuations are concentrated in closed surfaces.

4- The last model to be mentioned is the simplest one. It will be
discussed here in more detail because it furnished the starting
roots of the present discussion. This model is the ``homogeneous''
vacuum model, in which it is assumed that a magnetic field exist
in the vacuum \cite{Savv1}.

In the homogeneous vacuum field model, the existence of a constant
magnetic abelian field $H$ is assumed. A simple calculation in the
one loop approximation gives as result the following energy
density \cite{Shuryak}
\[
E\left(H\right) =\frac{H^2}2\left(1+\frac{bg^2}{16\pi ^2}\ln
\left(\frac H{\Lambda ^2}\right) \right).
\]
This formula predicts negative energy values for small values of
the field $ H $, so the usual perturbative ground state with $H=0$
is unstable with respect to the formation of a state with a non
vanishing field intensity \cite{Shuryak}
\[
H_{vac}=\Lambda ^2\exp \left(-\frac{16\pi ^2}{bg^2}-\frac
12\right),
\]
at which the energy $E\left(H\right)$ has a minimum.

With the use of this model an extensive number of physical
problems, related with the hadron structure, confinement, etc.
have been investigated. Nevertheless, after some time its intense
study was abandoned. The main reason were:

1. The perturbative relation giving $E_{vac}$ would be only valid
if the second order of the perturbative expansion is relatively
small.

2. The specific spatial and color directions of the magnetic field
break the now seemingly indispensable Lorentz and $SU(3)$
invariance of the ground state.

3. The magnetic moment of the vector particle (gluon) is such that
its energy in the presence of the field has a negative eigenvalue,
which also makes unstable the homogeneous magnetic field $H$.

Before presenting the objectives of the present work it should be
stressed that QCD quantization \cite{Faddeev-Popov} is realized in
the same way as that in QED, and it can be shown that QCD is
renormalizable. The quadratic field terms in the QCD Lagrangian
($L_{QCD}$), which depend on the quark and gluon fields, have the
same form that the ones corresponding to the electrons and photons
in QED. However, in connection with the interaction, there appears
a substantial difference due to the coupling of the gluon to
itself. In order to assure the unitarity of the quantum theory of
gauge fields, it was necessary to introduce fictitious particles
called the Faddeev-Popov ghosts, which carry color charge, behaves
as fermions (their fields anticommute) in spite of their boson
like propagation. These particles cancel out the contributions of
the non-physical gauge field degrees of freedom, and in physical
calculations only appear as internal lines of the Feynman
diagrams.

As it is well known, a perturbative expansion depends on the
initial conditions at $t\rightarrow \pm \infty $ or what is the
same on the states in which the expansion is based. The
perturbation theory at finite orders differs in attention to the
ground state selected, or from a functional point of view what
boundary conditions are chosen. The perturbation theory in QED
(PQED) is in excellent correspondence with the experimental facts.
In this theory the expansion is based on a perturbative vacuum
state that is the empty of the Fock space, excluding the presence
of fermion and boson particles. This is a radical simplification
of the exact perturbative ground state that should be a complex
combination of states on the Fock basis. Formally the expansion
around the Fock vacuum contains all the effects associated to the
exact vacuum, but it would require from infinite orders of the
expansion in the coupling constant for describing them. The rapid
convergence of the perturbative series in PQED indicates that the
higher excited states of the Fock basis expansion, in the real
vacuum, have a short life and a small influence on physical
observable. In QCD the color confinement indicates that the ground
state has a non-trivial structure, which in terms of a Fock
expansion could be represented with the formation of a gluon
``condensate''. Therefore, it should not be surprising that the
PQCD fails to describe the low energy physics where the propagator
of gluons could be affected by the presence of the condensate,
even under the validity of a modified perturbative expansion. Such
a perturbative condensate could generate all the effects over the
physical observable, which in the standard expansion could require
an infinite number of terms of the series.

In a previous work \cite{Cabo}, following the above ideas, the
construction of a modified perturbation theory for QCD was
implemented. This construction retained the main invariance of the
theory (the Lorentz and $SU(3)$ ones), and it was also able to
reproduce some of the main physical predictions of the
chromomagnetic field models. The central idea in that work was to
modify the perturbative expansion in such a way that the effects
of a gluon condensate could be incorporated. Such a modification
is needed to be searched through the connection of the interaction
on an alternative state in the Fock space designed to incorporate
the presence of the gluon condensate. It is not excluded that this
procedure could be also a crude approximation of the reality as in
the case in which the connection is done on the Fock vacuum (QED).
However, this procedure could produce a reasonable if not good
description of the low energy physics. If such is the case the low
and high-energy descriptions of QCD could be unified in a common
unified perturbation theory. In particular, in that previous work
\cite{Cabo} the results had the interesting outcome of producing a
non vanishing mean value for the relevant quantity $G^2$. In
addition the effective potential, in terms of the condensation
parameter at a first order approximation, showed a minimum at
non-vanishing values of that parameter. Therefore, the procedure
was able to reproduce at least some central predictions of the
chromomagnetic models and general QCD analysis.

The main objective of the present work is to search the
foundations of the mentioned perturbation theory. The concrete aim
is to find a physical state in the Fock space of the
non-interacting theory being able to generate that expansion. The
canonical quantization formalism for gauge fields, developed by
Kugo and Ojima is employed.

The exposition will be organized as follows: The Chapter 2 is
divided in three sections. In the first one a review of the former
work \cite{Cabo} is done, by also establishing the needs for the
present one and the objectives which are planned to be analyzed
and solved. In the second section the operational quantization
method for gauge fields developed by Kugo and Ojima is discussed.
Starting from it, in the third section it is exposed the ansatz
for the Fock space state that generates the desired form of the
perturbative expansion. The proof that the state satisfies the
physical state condition is also given in this section. The
Chapter 3 is divided in three sections. In the first one an
analysis for the general form of the generating functional in an
arbitrary ground state is made. In the second section it is shown
that the proposed state can generate the desired modification for
the gluon propagator by a proper selection of the parameters at
hand. In the third section the modification of the propagator for
the ghost particles is investigated, such propagator was not
modified in the work \cite{Cabo} and here this procedure is
justified as compatible within the present description. Finally,
two appendices are introduced for a detailed analysis of the most
elaborated parts in the calculation of transverse, longitudinal
and scalar modes contribution to the gluon propagator
modification.

\chapter{Ground State Ansatz}

The previous work \cite{Cabo} is reviewed, as motivation for the
present discussion, and the objectives for the present work
stated. It is also reviewed the canonical quantization method for
gauge fields developed by Kugo and Ojima (K.O.). Finally the QCD
modified vacuum state is proposed and it is shown that this state
satisfies the BRST physicality conditions imposed by the K.O.
formalism.

\section{Motivation}

In this section a review of a previous work \cite{Cabo} is made.
The main properties of this approach, as was mentioned in the
introduction, were:

a) The ability to produce a gluon-condensation parameter value
$\left\langle G^2\right\rangle $ directly in the first
approximation.

b) The prediction of a minimum of the effective action for
non-vanishing values of the condensation parameter.

The discussion in \cite{Cabo} opened the possibility of
reproducing some interesting physical implications of the early
chromomagnetic field models for the QCD vacuum \cite{Savv1,Savv2}
by also solving some of their main shortcoming: The breaking of
Lorentz and $SU\left(3\right)$ invariance. However, the discussion
in \cite{Cabo} had also a limitation; that is it was unknown if
the state that generated the proposed modification to the gluon
propagator was a physical state of the theory. This shortcoming,
could be expressed in the gauge parameter dependence of the
calculated gluon mass. Below it is reminded the main analysis in
\cite{Cabo}.

\newpage

The exposition was referred to the Euclidean space and the
followed conventions were used,
\begin{eqnarray*}
\nabla _\mu ^{ab} &=&\delta ^{ab}\partial _\mu +gf^{abc}A_\mu ^c,
\\ F_{\mu \nu }^a &=&\partial _\mu A_\nu ^a-\partial _\nu A_\mu
^a+f^{abc}A_\mu ^cA_\nu ^b,
\end{eqnarray*}
where $g$ is the coupling constant and $f^{abc}$ are the structure
constant of $SU(3)$.

The action for the problem, including the auxiliary sources for
all the fields was taken as,
\begin{eqnarray*}
S_T\left[ A,\overline{C},C\right] &=&\int d^4x\left\{ -\frac
14F_{\mu \nu }^aF_{\mu \nu }^a+\frac 1{2\alpha }\partial _\mu
A_\mu ^a\partial _\nu A_\nu ^a+\overline{C}^a\nabla _\mu
^{ab}\partial _\mu C^b\right. \\ &&\text{ \qquad \qquad \qquad
\qquad \qquad \qquad }\left. +J_\mu ^aA_\mu ^a+ \overline{\xi
}^aC^a+\overline{C}^a\xi ^a\right\},
\end{eqnarray*}
where $A_\mu,$ $\overline{C},C$ are the gauge and ghost fields and
$\alpha $ is the gauge fixing parameter \cite{Faddeev} for the
Lorentz gauge.

The generating functional for the Green functions was expressed in
the form
\[
Z_T\left[ J,\xi,\overline{\xi }\right] =\frac 1N\int D\left(
A,\overline{C},C\right) \exp \left\{ S_T\left[
A,\overline{C},C\right] \right\},
\]
which through the usual Legendre transformation led to the
effective action,
\begin{equation}
\Gamma \left[ \Phi \right] =\ln Z\left[ J\right] -J_i\Phi
_i,\text{ \quad with ~}\Phi _i=\frac{\delta \ln Z\left[ J\right]
}{\delta J_i}, \label{efec}
\end{equation}
$\Phi _i$ denoted the mean values of the fields, and the compact
notation of DeWitt \cite{Daemi},
\[
\Phi _i\equiv \left(A,\overline{C},C\right) ;\text{ \qquad
}J_i\equiv \left(J,\xi,\overline{\xi }\right),
\]
was used. In which $\Phi _i$ and $J_i$ indicate all the fields and
sources at a space-time point, respectively. Repeated indices
imply space-time integration as well as summation over all the
field types and over their Lorentz and color components.

The one-loop effective action and the corresponding ``quantum''
Lagrange equations, in the compact notation, were considered as,
\begin{eqnarray}
\Gamma \left[ \Phi \right] &=&S\left[ \Phi \right] +\frac 12\ln
DetD\left[ \Phi \right], \label{acc} \\ \Gamma _{,i}\left[ \Phi
\right] &=&S_{,i}\left[ \Phi \right] +\frac 12S_{,ikj}D_{kj}=-J_i,
\label{ecmo}
\end{eqnarray}
the functional derivatives were denoted by
\[
L_{,i}\left[ \Phi \right] =\frac{\delta L\left[ \Phi \right]
}{\delta \Phi _i }
\]
and the action defined by $S_T=S+J_i\Phi _i$.

The $\Phi$ dependent propagator $D$ was defined, as usual, through
\begin{equation}
D_{ij}=-S_{,ij}^{-1}\left[ \Phi \right], \label{D}
\end{equation}

After considering a null mean value for the vector field $\Phi$,
as requires the $SO(4)$ invariance, the propagator relation
(\ref{D}) took the form
\begin{equation}
D_{ij}=-S_{,ij}^{-1}\left[ 0\right].
\end{equation}

In this case the only non vanishing second derivatives of the
action were,
\begin{eqnarray}
\frac{\delta ^2S}{\delta A_\mu ^a\left(x\right) \delta A_\nu
^b\left(x^{\prime }\right) }\left[ 0\right] &=&\delta ^{ab}\left(
\partial _{x}^2\delta _{\mu \nu }-\left(1+\frac
1\alpha \right) \partial _\mu ^{x}\partial _\nu ^{x}\right) \delta
\left(x-x^{\prime }\right), \label{Sglu} \\ \frac{\delta
^2S}{\delta C^a\left(x\right) \delta \overline{C}^b\left(
x^{\prime }\right) }\left[ 0\right] &=&\delta ^{ab}\partial
_x^2\delta \left(x-x^{\prime }\right). \label{Sgho}
\end{eqnarray}

The gluon and ghost propagators are the inverse kernels of
(\ref{Sglu}) and (\ref{Sgho}). Here, the alternative for a
perturbative description of gluon condensation appeared. As
(\ref{Sglu}) consist of derivatives only, the inverse kernel of
the gluon propagator could include coordinate independent terms
reflecting a sort of gluon condensation. It should be noticed that
the propagator is a $SO\left(4\right)$ tensor (not a vector) then
a constant term in it does not led necessary to a breaking of the
$SO\left(4\right)$ invariance \cite{Cabo}. Accordingly with the
above remark gluon and ghost propagators were selected as
\begin{eqnarray}
D_{\mu \nu }^{ab}\left(x\right) &=&\int \frac{dp}{\left(2\pi
\right) ^4} \left[ C\delta ^{ab}\delta _{\mu \nu }\delta \left(
p\right) +\frac{\delta ^{ab}}{p^2}\left(\delta _{\mu \nu }-\left(
1+\alpha \right) \frac{p_\mu p_\nu }{p^2}\right) \right] \exp
\left(ipx\right), \label{Dglu} \\ D_G^{ab}\left(x\right) &=&\int
\frac{dp}{\left(2\pi \right) ^4}\frac{ \delta ^{ab}}{p^2}\exp
\left(ipx\right), \label{Dgho}
\end{eqnarray}
and it was checked that the equations of motion (\ref{ecmo}),
considering (\ref{Dglu}) and (\ref{Dgho}) and taking vanishing
gluon and ghost fields, were satisfied.

After that some implications of the modified gluon propagator, in
the standard perturbative calculations, were analyzed \cite{Cabo}.

The first interesting result obtained was the standard one loop
polarization tensor. It was modified by a massive term, depending
on the condensate parameter, with the form
\begin{equation}
m^2=\frac{3g^2}{\left(2\pi \right) ^4}C\left(1-\alpha \right).
\label{mass}
\end{equation}

This result had a dependence on the gauge parameter $\alpha$;
which as was mentioned above is one of the shortcomings of the
discussion \cite{Cabo} because it was unknown if this mass term
was generated by a non-physical vacuum state. In the present work
the idea is to solve this difficulty by explicitly constructing a
perturbative state leading to the considered form of the
propagator, but also satisfying the BRST physical state condition
in the non-interacting limit.

The mean value of the squared field intensity operator was also
calculated \cite{Cabo}, within the simplest approximation (the
tree approximation), with the use of the proposed propagator. That
is, it was evaluated the expression
\[
\langle 0\mid S_g\left[ \Phi \right] \mid 0\rangle \equiv \frac
1N\left[ \int D\left(\Phi \right) S_g\left[ \Phi \right] \exp
S_T\left[ \Phi \right] \right] _{J_i=0},
\]
with
\[
S_g\left[ \Phi \right] \equiv \int d^4x\left\{ -\frac 14F_{\mu \nu
}^a\left(x\right) F_{\mu \nu }^a\left(x\right) \right\},
\]
and the following result was obtained,
\begin{equation}
\langle 0\mid S_g\left[ \Phi \right] \mid 0\rangle
=-\frac{72g^2C^2}{\left(2\pi \right) ^8}\int d^4x.
\end{equation}

Then the mean value of $G^2$ took the form
\begin{equation}
G^2\equiv \langle 0\mid F_{\mu \nu }^a\left(x\right) F_{\mu \nu
}^a\left(x\right) \mid 0\rangle =\frac{288g^2C^2}{\left(2\pi
\right) ^8}. \label{G2}
\end{equation}

The substitution of Eq. (\ref{G2}) in Eq. (\ref{mass}) gave a
rough estimate of the gluon mass. It was selected a particular
value of $\alpha =0$ and assumed the more or less accepted value
of $g^2G^2$ in the physical vacuum
\begin{equation}
g^2G^2\cong 0.5\left(\frac{GeV}{c^2}\right) ^4,
\end{equation}
then the estimated value of the gluon mass became
\begin{equation}
m=0.35\frac{GeV}{c^2}.
\end{equation}

Finally, an evaluation for the contribution to the effective
potential of all the one-loop graphs, having only mass term
insertions in the polarization tensor, was done. The result, in
terms of $G^2$ (\ref{G2}), turned to be of the form \cite{Cabo},
\begin{equation}
V\left(G^2\right) =\frac{G^2}4+\frac 3{16\pi
^2}g^2\frac{G^2}{32}\ln \frac{ g^2G^2}{\mu ^4}, \label{VG}
\end{equation}
where $\mu $ is the dimensional parameter included by the
renormalization procedure.

As it can be noticed in (\ref{VG}), the effective potential
indicates the spontaneous generation of a $G^2$ condensate from
the usual perturbative vacuum $\left(G^2=0\right) $. This occurred
in close analogy with the chromomagnetic fields.

Then from the reviewed functional treatment, there are some
interesting features that allow believing that the above procedure
could describe relevant phenomena of the low energy region through
a perturbative expansion. However some questions needed to be
answered and taken as objectives of the present work are:

1- To determine under what conditions the new gluon propagator
(\ref{Dglu}) corresponds to a modified vacuum satisfying the
physical state condition. This could also help in the
understanding of the $\alpha$ dependence in the gluonic mass term.

2- To investigate the form of the ghost propagator in the modified
vacuum state, because in the previous work \cite{Cabo} it was
taken the as same of the usual perturbative theory.

\section{Operational Quantization Formalism}

As it is well known the non-abelian character of Yang-Mills fields
determines the asymptotic freedom property, and the
quark-confinement problem of QCD. This character simultaneously
makes difficult the quantization of such theories. The first
approach to this quantization was made by Faddeev and Popov in the
path integral formalism \cite{Faddeev-Popov}, with the resulting
correct Feynman rules including the Faddeev-Popov ghost fields and
the renormalizability of the theory. But this approach has the
problem of the absence of notions about the state vector space and
the Heisemberg operators. In this case due the non-abelian
character of the theory it is not possible to use the operators
formalism developed by Gupta-Bleuler \cite{Gupta} or the more
general Nakanishi-Lautrup version \cite{Nakanishi}, which can be
used only for the abelian case. This situation occurs because de
S-Matrix calculated with those procedures is not unitary in the
non abelian case, as it was first mentioned by Feynman
\cite{Feynman}.

In the present work the operator formalism developed by T. Kugo
and I. Ojima \cite{Kugo}, for the first consistent quantization of
the Yang-Mills fields, is considered. This formulation uses the
Lagrangian invariance under a global symmetry operation called the
BRST transformation \cite{BRST}. In the following a brief review
of the K.O. work is done and the following conventions are used.

Let $G$ be a compact Lie group, and $\Lambda$ any matrix in the
adjoin representation of its associated Lie Algebra. The matrix
$\Lambda$ can be represented as a linear combination of the form
\[
\Lambda =\Lambda ^aT^a,
\]
were $T^a$ are the generators $(a=1,...,$Dim$G=n)$, which can be
chosen as Hermitian ones and satisfying
\[
\left[ T^a,T^b\right] =if^{abc}T^c.
\]

The field variations under infinitesimal gauge transformations are
given by
\begin{eqnarray*}
\delta _\Lambda A_\mu ^a\left(x\right) &=&\partial _\mu \Lambda
^a\left(x\right) +gf^{acb}A_\mu ^c\left(x\right) \Lambda ^b\left(
x\right) =D_\mu ^{ab}\left(x\right) \Lambda ^b, \\ D_\mu
^{ab}\left(x\right) &=&\partial _\mu \delta ^{ab}+gf^{acb}A_\mu
^c\left(x\right).
\end{eqnarray*}

The metric $g_{\mu \nu }$ is taken in the convention
\[
g_{00}=-g_{ii}=1\qquad \text{for}\quad i=1,2,3.
\]

The complete G.D. Lagrangian to be considered is the one employed
in the operator quantization approach \cite{OjimaTex}. Its
explicit form is given by
\begin{eqnarray}
\mathcal{L} &=&\mathcal{L}_{YM}+\mathcal{L}_{GF}+\mathcal{L}_{FP}
\label{Lag} \\ \mathcal{L}_{YM} &=&-\frac 14F_{\mu \nu }^a\left(
x\right) F^{\mu \nu,a}\left(x\right), \label{YM} \\
\mathcal{L}_{GF} &=&-\partial ^\mu B^a\left(x\right) A_\mu
^a\left(x\right) +\frac \alpha 2B^a\left(x\right) B^a\left(
x\right), \label{GF} \\ \mathcal{L}_{FP} &=&-i\partial ^\mu
\overline{c}^a\left(x\right) D_\mu ^{ab}\left(x\right) c^b\left(
x\right), \label{FP}
\end{eqnarray}
where field intensity is
\[
F_{\mu \nu }^a\left(x\right) =\partial _\mu A_\nu ^a\left(
x\right) -\partial _\nu A_\mu ^a\left(x\right) +gf^{abc}A_\mu
^b\left(x\right) A_\nu ^c\left(x\right).
\]

Relation (\ref{YM}) defines the standard Yang-Mills Lagrangian,
Eq. (\ref{GF}) defines the gauge fixing term which can be also
rewritten in the form
\[
\mathcal{L}_{GF}=-\frac 1{2\alpha }\left(\partial ^\mu A_\mu
^a\left(x\right) \right) ^2+\frac \alpha 2\left(B^a\left(x\right)
+\frac 1\alpha
\partial ^\mu A_\mu ^a\left(x\right) \right) ^2-\partial ^\mu \left(
B^a\left(x\right) A_\mu ^a\left(x\right) \right),
\]
equivalent to the more familiar $-\frac 1{2\alpha }\left(\partial
^\mu A_\mu ^a\left(x\right) \right) ^2$, at the equations of
motion level \cite{Faddeev} and Feynman diagram expansion.

Finally, Eq. (\ref{FP}) describes the non-physical Faddeev-Popov
ghost sector. The definition for such fields in the Kugo and Ojima
(K.O.) approach is satisfying
\[
\overline{c}^{\dagger }=\overline{c},\text{ \qquad }c^{\dagger
}=c.
\]

That is, the ghost fields are Hermitian. In the Faddeev-Popov
formalism \cite{Faddeev} they satisfy
\[
C^{\dagger }=\overline{C},\text{ \qquad }\overline{C}^{\dagger
}=C.
\]

However, a simple change of variables is able to transform between
the ghost fields satisfying both kind of conjugation conditions.
The selected conjugation properties, for this sector, allowed Kugo
and Ojima to solve various formal problems existing for the
application of the BRST operator quantization method to QCD, for
example the hermiticity of the Lagrangian, which guarantees the
unitarity of the S-Matrix.

The physical state conditions in the BRST procedure
\cite{OjimaTex} are given by
\begin{eqnarray}
&&Q_B\mid phys\rangle =0, \nonumber \\ &&Q_C\mid phys\rangle =0,
\end{eqnarray}
where
\[
Q_B=\int d^3x\left[ B^a\left(x\right) \overleftrightarrow{\partial
_0} c^a\left(x\right) +gB^a\left(x\right) f^{abc}A_0^b\left(
x\right) c^c\left(x\right) +\frac i2g\partial _0\left(
\overline{c}^a\right) f^{abc}c^b\left(x\right) c^c\left(x\right)
\right],
\]
with
\[
f\left(x\right) \overleftrightarrow{\partial _0}g\left(x\right)
\equiv f\left(x\right) \partial _0g\left(x\right) -\partial
_0\left(f\left(x\right) \right) g\left(x\right).
\]

The BRST charge is conserved as a consequence of the BRST symmetry
of the Lagrangian (\ref{Lag}).

The also conserved charge $Q_C$ is given by
\[
Q_C=i\int d^3x\left[ \overline{c}^a\left(x\right)
\overleftrightarrow{
\partial _0}c^a\left(x\right) +g\overline{c}^a\left(x\right)
f^{abc}A_0^b\left(x\right) c^c\left(x\right) \right],
\]
its conservation comes from the Noether theorem, due to the
Lagrangian invariance (\ref{Lag}) under the phase transformation
$c\rightarrow e^\theta c,\ \overline{c}\rightarrow e^{-\theta
}\overline{c}$. This charge defines the so called ``ghost number''
as the difference between the number of ghost $c$ and
$\overline{c}$.

The analysis here is restricted to the Yang-Mills Theory without
spontaneous breaking of the gauge symmetry. The quantization for
the theory defined by the Lagrangian (\ref{Lag}), considering the
interacting free limit $g\rightarrow 0$, leads to the following
commutation relations between the free fields,
\begin{eqnarray}
\left[ A_\mu ^a\left(x\right),A_\nu ^b\left(y\right) \right]
&=&\delta ^{ab}\left(-ig_{\mu \nu }D\left(x-y\right) +i\left(
1-\alpha \right) \partial _\mu \partial _\nu E\left(x-y\right)
\right), \nonumber \\ \left[ A_\mu ^a\left(x\right),B^b\left(
y\right) \right] &=&\delta ^{ab}\left(-i\partial _\mu D\left(
x-y\right) \right), \nonumber \\ \left[ B^a\left(
x\right),B^b\left(y\right) \right] &=&\left\{ \overline{c}
^a\left(x\right),\overline{c}^b\left(y\right) \right\} =\left\{
c^a\left(x\right),c^b\left(y\right) \right\} =0, \nonumber \\
\left\{ c^a\left(x\right),\overline{c}^b\left(y\right) \right\}
&=&-D\left(x-y\right), \label{com}
\end{eqnarray}
The $E$ functions are defined by \cite{OjimaTex}
\[
E_{\left(.\right) }\left(x\right) =\frac 12\left(\nabla ^2\right)
^{-1}\left(x_0\partial ^0-1\right) D_{\left(.\right) }\left(
x\right).
\]

The equations of motion for the non-interacting fields take the
simple form
\begin{eqnarray}
\Box A_\mu ^a\left(x\right) -\left(1-\alpha \right) \partial _\mu
B^a\left(x\right) &=&0, \\ \partial ^\mu A_\mu ^a\left(x\right)
+\alpha B^a\left(x\right) &=&0, \label{liga1} \\ \Box B^a\left(
x\right) =\Box c^a\left(x\right) =\Box \overline{c} ^a\left(
x\right) &=&0.
\end{eqnarray}

This equations can be solved for an arbitrary values of the
$\alpha$ parameter. However, the discussion will be restricted to
the case $\alpha =1$ which corresponds to the situation in which
all the gluon components satisfy the D'Alambert equation. This
selection, as considered in the framework of the usual
perturbative expansion, implies that you are not able to check the
$\alpha$ independence of the physical quantities. This
simplification is a necessary requirement. In the present
discussion, the aim is to construct a perturbative state that
satisfies the BRST physical state condition, in order to connect
adiabatically the interaction. Then, the physical character of all
the prediction will follow whenever the former assumption that
adiabatic connection do not take the state out of the physical
subspace at any intermediate state. The consideration of different
values of $\alpha $, would be also a convenient recourse for
checking the $\alpha$ independent perturbative expansion. However,
at this stage it is preferred to delay this more technical issue
for future work.

In that way the field equations for the $\alpha =1$ are
\begin{eqnarray}
\Box A_\mu ^a\left(x\right)=\Box B^a\left(x\right) =\Box c^a\left(
x\right) =\Box \overline{c} ^a\left(x\right) &=&0, \label{movi1}
\\
\partial ^\mu A_\mu ^a\left(x\right) +B^a\left(x\right) &=&0.
\label{movi2}
\end{eqnarray}

The solutions of the set (\ref{movi1}), (\ref{movi2}) can be
written as
\begin{eqnarray}
A_\mu ^a\left(x\right) &=&\sum\limits_{\vec{k},\sigma }\left(
A_{\vec{k},\sigma }^af_{k,\mu }^\sigma \left(x\right)
+A_{\vec{k},\sigma }^{a+}f_{k,\mu }^{\sigma *}\left(x\right)
\right), \nonumber \\ B^a\left(x\right)
&=&\sum\limits_{\vec{k}}\left(B_{\vec{k}}^ag_k\left(x\right)
+B_{\vec{k}}^{a+}g_k^{*}\left(x\right) \right), \nonumber \\
c^a\left(x\right) &=&\sum\limits_{\vec{k}}\left(
c_{\vec{k}}^ag_k\left(x\right) +c_{\vec{k}}^{a+}g_k^{*}\left(
x\right) \right), \nonumber \\ \overline{c}^a\left(x\right)
&=&\sum\limits_{\vec{k}}\left(\overline{c}_{ \vec{k}}^ag_k\left(
x\right) +\overline{c}_{\vec{k}}^{a+}g_k^{*}\left(x\right)
\right).
\end{eqnarray}
The wave packets system, for non-massive scalar and vector fields,
are taken in the form
\begin{eqnarray}
g_k\left(x\right) &=&\frac 1{\sqrt{2Vk_0}}\exp \left(-ikx\right),
\nonumber \\ f_{k,\mu }^\sigma \left(x\right) &=&\frac
1{\sqrt{2Vk_0}}\epsilon _\mu ^\sigma \left(k\right) \exp \left(
-ikx\right). \label{pol}
\end{eqnarray}

The polarization vectors, in Eq. (\ref{pol}) are defined by
\[
\vec{k}\cdot \vec{\epsilon}_\sigma \left(k\right) =0,\ \epsilon
_\sigma ^0\left(k\right) =0,
\]
and satisfy
\[
\vec{\epsilon}_\sigma \left(k\right) \cdot \vec{\epsilon}_\tau
\left(k\right) =\delta _{\sigma \tau },
\]
where $\sigma,\tau =1,2$ are the transverse modes. For the
longitudinal $L$ and scalar $S$ modes the definitions are
\begin{eqnarray*}
\epsilon _L^\mu \left(k\right) &=&-ik^\mu =-i\left(\left|
\vec{k}\right|, \vec{k}\right),\ \epsilon _L^{\mu *}\left(
k\right) =-\epsilon _L^\mu \left(k\right), \\ \epsilon _S^\mu
\left(k\right) &=&-i\frac{\overline{k}^\mu }{2\left| \vec{k}
\right| ^2}=\frac{-i\left(\left| \vec{k}\right|,-\vec{k}\right)
}{2\left| \vec{k}\right| ^2},\ \epsilon _S^{\mu *}\left(k\right)
=-\epsilon _S^\mu \left(k\right),
\end{eqnarray*}
and satisfy
\begin{eqnarray*}
\epsilon _L^{\mu *}\left(k\right) \cdot \epsilon _{L,\mu }\left(
k\right) &=&\epsilon _S^{\mu *}\left(k\right) \cdot \epsilon
_{S,\mu }\left(k\right) =0, \\ \epsilon _L^{\mu *}\left(k\right)
\cdot \epsilon _{S,\mu }\left(k\right) &=&1.
\end{eqnarray*}

The scalar product of the defined polarizations define the metric
matrix
\[
\widetilde{\eta }_{\sigma \tau }=\epsilon _\sigma ^{\mu
*}\left(k\right) \cdot \epsilon _{\tau,\mu }\left(k\right)\equiv
\left(
\begin{array}{cccc}
-1 & 0 & 0 & 0 \\
0 & -1 & 0 & 0 \\
0 & 0 & 0 & 1 \\
0 & 0 & 1 & 0
\end{array}
\right).
\]

Now it is possible to introduce the contravariant (in the
polarization index) polarizations
\[
\epsilon ^{\sigma,\mu }\left(k\right)
=\sum\limits_{1,2,L,S}\widetilde{ \eta }^{\sigma \tau }\cdot
\epsilon _\tau ^\mu \left(k\right),
\]
satisfying
\[
\sum\limits_\sigma \epsilon ^{\sigma,\mu }\left(k\right) \cdot
\epsilon _\sigma ^{\nu *}\left(k\right) =\sum\limits_{\sigma,\tau
}\widetilde{\eta } ^{\sigma \tau }\cdot \epsilon _\tau ^\mu \left(
k\right) \cdot \epsilon _\sigma ^{\nu *}\left(k\right) =g^{\mu \nu
}
\]
and
\begin{eqnarray*}
\epsilon ^{\sigma,\mu }\left(k\right) \cdot \epsilon _\mu ^{\tau
*}\left(k\right) &=&\widetilde{\eta }^{\sigma \tau }, \\
\widetilde{\eta }^{\sigma \tau ^{\prime }}\cdot \widetilde{\eta
}_{\tau ^{\prime }\tau } &=&\delta _\tau ^\sigma.
\end{eqnarray*}

After that, it follows for the vector functions
\[
\sum\limits_{\vec{k},\sigma }f_{k,\sigma }^\mu \left(x\right)
\cdot f_k^{\sigma,\nu *}\left(y\right) =g^{\mu \nu }D_{+}\left(
x-y\right).
\]

As it can be seen from (\ref{movi2}) the $A_{\vec{k},\sigma }^a$
and $B_{\vec{k}}^a$ modes are not all independent. Indeed, it
follows from (\ref {movi2}) that
\[
B_{\vec{k}}^a=A_{\vec{k}}^{S,a}=A_{\vec{k},L}^a.
\]

Then excluding the scalar mode, the free Heisemberg fields
expansion takes the form
\begin{equation}
A_\mu ^a\left(x\right) =\sum\limits_{\vec{k}}\left(
\sum\limits_{\sigma =1,2}A_{\vec{k},\sigma }^af_{k,\mu }^\sigma
\left(x\right) +A_{\vec{k} }^{L,a}f_{k,L,\mu }\left(x\right)
+B_{\vec{k}}^af_{k,S,\mu }\left(x\right) \right)+h.c.,
\end{equation}
where $h.c.$ represents the Hermitian conjugate of the first term.

\newpage

In order to satisfy the commutations relations (\ref{com}) the
creation and annihilation operator, associated to the Fourier
components of the field, should obey
\begin{eqnarray}
\left[ A_{\vec{k},\sigma }^a,A_{\vec{k}^{\prime },\sigma ^{\prime
}}^{a^{\prime }+}\right] &=&-\delta ^{aa^{\prime }}\delta
_{\vec{k}\vec{k} ^{\prime }}\eta _{\sigma \sigma ^{\prime }},
\nonumber \\ \left\{ c_{\vec{k}}^a,\overline{c}_{\vec{k}^{\prime
}}^{a^{\prime }+}\right\} &=&i\delta ^{aa^{\prime }}\delta
_{\vec{k}\vec{k}^{\prime }}, \nonumber \\ \left\{
\overline{c}_{\vec{k}}^a,c_{\vec{k}^{\prime }}^{a^{\prime
}+}\right\} &=&-i\delta ^{aa^{\prime }}\delta
_{\vec{k}\vec{k}^{\prime }}
\end{eqnarray}
and all the other vanish.

In a symbolic matrix form these relations can be arranged as
follows
\begin{equation}
\begin{array}{cccccc}
& A_{\vec{k}^{\prime },\sigma ^{\prime }}^{a^{\prime }+} &
A_{\vec{k} ^{\prime }}^{L,a^{\prime }+} & B_{\vec{k}^{\prime
}}^{a^{\prime }+} & c_{ \vec{k}^{\prime }}^{a+} &
\overline{c}_{\vec{k}^{\prime }}^{a+} \\ A_{\vec{k},\sigma }^a &
\delta ^{aa^{\prime }}\delta _{\vec{k}\vec{k} ^{\prime }}\delta
_{\sigma \sigma ^{\prime }} & 0 & 0 & 0 & 0 \\ A_{\vec{k}}^{L,a} &
0 & 0 & -\delta ^{aa^{\prime }}\delta _{\vec{k}\vec{k} ^{\prime }}
& 0 & 0 \\ B_{\vec{k}}^a & 0 & -\delta ^{aa^{\prime }}\delta
_{\vec{k}\vec{k}^{\prime }} & 0 & 0 & 0 \\ c_{\vec{k}}^a & 0 & 0 &
0 & 0 & i\delta ^{aa^{\prime }}\delta _{\vec{k}\vec{k }^{\prime }}
\\ \overline{c}_{\vec{k}}^a & 0 & 0 & 0 & -i\delta ^{aa^{\prime
}}\delta _{\vec{ k}\vec{k}^{\prime }} & 0
\end{array}
\label{commu}
\end{equation}

The above commutation rules and equation of motions define the
quantized non-interacting limit of G.D. Then, it is possible now
to define the alternative interacting free ground state to be
considered for the adiabatic connection of the interaction. As
discussed before, the expectation is that the physics of the
perturbation theory to be developed will be able to furnish good
description of some low energy physical effects.

It is interesting to comment now that one of the first tasks
proposed for the present work was to construct a state, in quantum
electrodynamics, able to generate a modification for the photon
propagator similar to the one proposed in \cite{Cabo} for gluons.
It was used the quantification operator method developed by Gupta
and Bleuler (GB), however was impossible to find any state
generating this covariant propagator modification and satisfying
the physical state condition imposed by this formalism.

In the GB formalism the physical state condition is given by
\[
\partial ^\mu A_\mu ^{+}\left(x\right) \mid \Phi \rangle =0,
\]
or in terms of the annihilation operators \cite{Sokolov}, by
\[
k_0\left(A_{\vec{k},3}-A_{\vec{k},0}\right) \mid \Phi \rangle =0.
\]

The more general state satisfying this condition is
\cite{GuptaTex}
\[
\mid \Phi \rangle =\sum\limits_{m,n_{1,}n_2}B_{n_{1,}n_2,m}\mid
\Phi \left(n_{1,}n_2,m\right) \rangle,
\]
with
\[
\mid \Phi \left(n_{1,}n_2,m\right) \rangle =\left(m!\right)
^{-\frac 12}\left(A_{\vec{k},3}^{+}-A_{\vec{k},0}^{+}\right)
^m\left(n_1!n_2!\right) ^{-\frac 12}\left(
A_{\vec{k},1}^{+}\right) ^{n_1}\left(A_{ \vec{k},2}^{+}\right)
^{n_2}\mid 0\rangle,
\]
where $B_{n_{1,}n_2,m}$ are arbitrary constants. This general form
of the state is the one that disabled to find a covariant
modification to the propagator.

\section{The alternative vacuum state}

In the present section the construction of a relativistic
invariant ground state in the non-interacting limit of QCD is
considered. It will be required that the proposed state satisfies
the BRST physical state conditions. Then this state will have an
opportunity to furnish the gluodynamics ground state under the
adiabatic connection of the interaction.

After beginning to work in the K.O. formalism some indications
were found, that the appropriate state obeying the physical state
conditions in this procedure, and with possibilities for
generating the modification to the gluon propagator proposed in
the previous work, could have the general structure
\begin{equation}
\mid \phi \rangle =\exp \sum\limits_a\left(\sum\limits_{\sigma
=1,2}\frac 12C_\sigma \left(\left| \vec{p}\right| \right)
A_{\vec{p},\sigma }^{a+}A_{ \vec{p},\sigma }^{a+}+C_3\left(\left|
\vec{p}\right| \right) \left(B_{\vec{
p}}^{a+}A_{\vec{p}}^{L,a+}+i\overline{c}_{\vec{p}}^{a+}c_{\vec{p}
}^{a+}\right) \right) \mid 0\rangle, \label{Vacuum}
\end{equation}
where $\vec{p}$ is an auxiliary momentum chosen as one of the few
smallest momenta of the quantized theory in a finite volume $V$.
This value will be taken later in the limit $V\rightarrow \infty $
for recovering Lorentz invariance. From here the sum on the color
index $a$ will be explicit. The parameters $C_i\left(\left|
\vec{p}\right| \right)$, $i=1,2,3,$ will be fixed below from the
condition that the free propagator associated to a state
satisfying the BRST physical state condition, coincides with the
one proposed in the previous work \cite{Cabo}. The solution of
this problem, then would give a more solid foundation to the
physical implications of the discussion in that work.

It should also be noticed that the state defined by Eq.
(\ref{Vacuum}) has some similarity with the coherent states
\cite{Itzykson}. However, in the present case, the creation
operators appear in squares. Thus, the argument of the exponential
creates pairs of physical and non-physical particles. An important
property of this function is that its construction in terms of
creation operator pairs determines that the mean value of an odd
number of field operators vanishes. This at variance with the
standard coherent state in which the mean values of the fields are
nonzero. The vanishing of the mean fields is a property in common
with the standard perturbative vacuum, in which Lorentz invariance
could be broken by a non-vanishing expectation value of a 4-vector
the gauge field. It should be also stressed that this state as
formed by the superposition of gluons state pairs suggests a
connection with some recent works in the literature that consider
the formation gluons pairs due to the color interactions.

Now it is checked that the state (\ref{Vacuum}) satisfies the BRST
physical state conditions
\begin{eqnarray}
&&Q_B\mid \Phi \rangle =0, \nonumber \\ &&Q_C\mid \Phi \rangle =0.
\end{eqnarray}

The expressions for these charges in the interaction free limit
\cite{OjimaTex} are
\begin{eqnarray}
&&Q_B=i\sum\limits_{\vec{k},a}\left(
c_{\vec{k}}^{a+}B_{\vec{k}}^a-B_{\vec{k} }^{a+\
}c_{\vec{k}}^a\right), \nonumber \\
&&Q_C=i\sum\limits_{\vec{k},a}\left(
\overline{c}_{\vec{k}}^{a+}c_{\vec{k}
}^a+c_{\vec{k}}^{a+}\overline{c}_{\vec{k}}^a\right).
\end{eqnarray}

Considering first the action of $Q_B$ on the proposed state,
{\setlength\arraycolsep{0.5pt}
\begin{eqnarray}
&&Q_B\mid \Phi \rangle =i\exp \left\{ \sum\limits_{\sigma,a}\frac
12C_\sigma \left(\left| \vec{p}\right| \right) A_{\vec{p},\sigma
}^{a+}A_{\vec{p},\sigma }^{a+}\right\} \times \nonumber \\
&&\times \left(\exp \left\{ \sum\limits_aC_3\left(\left|
\vec{p}\right| \right)
i\overline{c}_{\vec{p}}^{a+}c_{\vec{p}}^{a+}\right\} \sum\limits_{
\vec{k},b}c_{\vec{k}}^{b+}B_{\vec{k}}^b\exp \left\{
\sum\limits_aC_3\left(\left| \vec{p}\right| \right)
B_{\vec{p}}^{a+}A_{\vec{p}}^{L,a+}\right\} \right. \\ &&-\left.
\exp \left\{ \sum\limits_aC_3\left(\left| \vec{p}\right| \right)
B_{\vec{p}}^{a+}A_{\vec{p}}^{L,a+}\right\}
\sum\limits_{\vec{k},b}B_{\vec{k} }^{b+}c_{\vec{k}}^b\exp \left\{
\sum\limits_aC_3\left(\left| \vec{p}\right| \right)
i\overline{c}_{\vec{p}}^{a+}c_{\vec{p}}^{a+}\right\} \right) \mid
0\rangle =0, \nonumber
\end{eqnarray}
}where the identity
\begin{equation}
\left[ B_{\vec{k}}^b,\exp \sum\limits_aC_3\left(\left|
\vec{p}\right| \right) B_{\vec{p}}^{a+}A_{\vec{p}}^{L,a+}\right]
=-C_3\left(\left| \vec{p} \right| \right) B_{\vec{p}}^{b+}\delta
_{\vec{k},\vec{p}}\exp \sum\limits_aC_3\left(\left| \vec{p}\right|
\right) B_{\vec{p}}^{a+}A_{\vec{ p}}^{L,a+}, \label{ident1}
\end{equation}
was used.

For the action of $Q_C$ on the considered state it follows
{\setlength\arraycolsep{0.1pt}
\begin{eqnarray}
&&Q_C\mid \Phi \rangle =i\exp \left\{ \sum\limits_{\sigma,a}\frac
12C_\sigma \left(\left| \vec{p}\right| \right)A_{\vec{p},\sigma
}^{a+}A_{\vec{p},\sigma }^{a+}+\sum\limits_aC_3\left(\left|
\vec{p}\right| \right) B_{\vec{p} }^{a+}A_{\vec{p}}^{L,a+}\right\}
\\ &&\times \left[ \sum\limits_{\vec{k},b}
\overline{c}_{\vec{k}}^{b+}c_{\vec{k} }^b\left(
1+\sum\limits_aiC_3\left(\left| \vec{p}\right| \right) \overline{
c }_{\vec{p}}^{a+}c_{\vec{p}}^{a+}\right)
+\sum\limits_{\vec{k},b}c_{\vec{k}
}^{b+}\overline{c}_{\vec{k}}^b\left(1+\sum\limits_aiC_3\left(
\left| \vec{p} \right| \right)
\overline{c}_{\vec{p}}^{a+}c_{\vec{p}}^{a+}\right) \right] \mid
0\rangle =0 \nonumber
\end{eqnarray}
}which vanishes due to the commutation rules of the ghost
operators (\ref {commu}).

Next, the evaluation of norm of the proposed state is considered,
which due to the commutation properties of the operator can be
written as

\begin{eqnarray}
\langle \Phi \mid \Phi \rangle =\prod\limits_{a=1,..,8}
&\prod\limits_{\sigma =1,2}&\langle 0\mid \exp \left\{ \frac
12C_\sigma ^{*}\left(\left| \vec{p}\right| \right)
A_{\vec{p},\sigma }^aA_{\vec{p},\sigma }^a\right\} \exp \left\{
\frac 12C_\sigma \left(\left| \vec{p} \right| \right)
A_{\vec{p},\sigma }^{a+}A_{\vec{p},\sigma }^{a+}\right\} \mid
0\rangle \nonumber \\ &\times &\langle 0\mid \exp \left\{
C_3^{*}\left(\left| \vec{p}\right| \right)
A_{\vec{p}}^{L,a}B_{\vec{p}}^a\right\} \exp \left\{ C_3\left(
\left| \vec{p}\right| \right)
B_{\vec{p}}^{a+}A_{\vec{p}}^{L,a+}\right\} \mid 0\rangle \nonumber
\\ &\times &\langle 0\mid \left(1-iC_3^{*}\left(\left|
\vec{p}\right| \right)
c_{\vec{p}}^a\overline{c}_{\vec{p}}^a\right) \left(1+iC_3\left(
\left| \vec{ p}\right| \right)
\overline{c}_{\vec{p}}^{a+}c_{\vec{p}}^{a+}\right) \mid 0\rangle.
\end{eqnarray}

For the product of the factors associated with transverse modes
and the eight values of the color index, after expanding the
exponential in series, it follows that

\begin{eqnarray}
&&\left[ \langle 0\mid \exp \left\{ \frac 12C_\sigma ^{*}\left(
\left| \vec{p }\right| \right) A_{\vec{p},\sigma
}^aA_{\vec{p},\sigma }^a\right\} \exp \left\{ \frac 12C_\sigma
\left(\left| \vec{p}\right| \right) A_{\vec{p},\sigma
}^{a+}A_{\vec{p},\sigma }^{a+}\right\} \mid 0\rangle \right] ^8
\nonumber \\ &&=\left[ \langle 0\mid \sum\limits_{m=0}^\infty
\left| \frac 12C_\sigma \left(\left| \vec{p}\right| \right)
\right| ^{2m}\frac{\left(A_{\vec{p},\sigma }^a\right) ^{2m}\left(
A_{\vec{p},\sigma }^{a+}\right) ^{2m}}{\left(m!\right) ^2}\mid
0\rangle \right] ^8 \nonumber \\ &&=\left[
\sum\limits_{m=0}^\infty \left| \frac 12C_\sigma \left(\left|
\vec{p}\right| \right) \right| ^{2m}\frac{\left(2m\right)
!}{\left(m!\right) ^2}\right] ^8, \label{normT}
\end{eqnarray}
where the identity
\[
\langle 0\mid \left(A_{\vec{p},\sigma }^a\right) ^{2m}\left(
A_{\vec{p},\sigma }^{a+}\right) ^{2m}\mid 0\rangle =\left(
2m\right) !,
\]
was used.

The factors linked with the scalar and longitudinal modes can be
transformed as follows
\begin{eqnarray}
&&\left[ \langle 0\mid \exp \left\{ C_3^{*}\left(\left|
\vec{p}\right| \right) A_{\vec{p}}^{L,a}B_{\vec{p}}^a\right\} \exp
\left\{ C_3\left(\left| \vec{p}\right| \right)
B_{\vec{p}}^{a+}A_{\vec{p}}^{L,a+}\right\} \mid 0\rangle \right]
^8 \nonumber \\ &&=\left[ \langle 0\mid \sum\limits_{m=0}^\infty
\left| C_3\left(\left| \vec{p}\right| \right) \right|
^{2m}\frac{\left(A_{\vec{p}}^{L,a}B_{\vec{p} }^a\right) ^m\left(
B_{\vec{p}}^{a+}A_{\vec{p}}^{L,a+}\right) ^m}{\left(m!\right)
^2}\mid 0\rangle \right] ^8 \nonumber \\ &&=\left[
\sum\limits_{m=0}^\infty \left| C_3\left(\left| \vec{p}\right|
\right) \right| ^{2m}\right] ^8=\left[ \frac 1{\left(1-\left|
C_3\left(\left| \vec{p}\right| \right) \right| ^2\right) }\right]
^8\text{ for}\quad \left| C_3\left(\left| \vec{p}\right| \right)
\right| <1, \label{normLS}
\end{eqnarray}
\noindent in which the identity
\[
\langle 0\mid \left(A_{\vec{p}}^{L,a}B_{\vec{p}}^a\right) ^m\left(
B_{\vec{p }}^{a+}A_{\vec{p}}^{L,a+}\right) ^m\mid 0\rangle =\left(
m!\right) ^2,
\]
was employed.

Finally the factor connected with the ghost fields can be
calculated as follows

\begin{eqnarray}
&&\left[ \langle 0\mid \left(1-iC_3^{*}\left(\left| \vec{p}\right|
\right) c_{\vec{p}}^a\overline{c}_{\vec{p}}^a\right) \left(
1+iC_3\left(\left| \vec{ p}\right| \right)
\overline{c}_{\vec{p}}^{a+}c_{\vec{p}}^{a+}\right) \mid 0\rangle
\right] ^8 \nonumber \\ &&=\left[ 1+\left| C_3\left(\left|
\vec{p}\right| \right) \right| ^2\langle 0\mid
c_{\vec{p}}^a\overline{c}_{\vec{p}}^a\overline{c}_{\vec{p}}^{a+}c_{
\vec{p}}^{a+}\mid 0\rangle \right] =\left[ 1-\left| C_3\left(
\left| \vec{p} \right| \right) \right| ^2\right] ^8. \label{normG}
\end{eqnarray}

After substituting all the calculated factors, the norm of the
state can be written as

\begin{equation}
N=\langle \Phi \mid \Phi \rangle =\prod\limits_{\sigma =1,2}\left[
\sum\limits_{m=0}^\infty \left| C_\sigma \left(\left|
\vec{p}\right| \right) \right| ^{2m}\frac{\left(2m\right)
!}{\left(m!\right) ^2}\right] ^8.
\end{equation}

Therefore, it is possible to define the normalized state
\begin{equation}
\mid \widetilde{\Phi }\rangle =\frac 1{\sqrt{N}}\mid \Phi \rangle.
\end{equation}
Note that, as it should be expected, the norm is not dependent on
the $ C_3\left(\left| \vec{p}\right| \right) $ parameter which
defines the non-physical particle operators entering in the
definition of the proposed vacuum state.

\chapter{Propagator Modifications}

The general form for generating functionals and propagators, for
boson and fermion particles in an arbitrary vacuum state, are
analyzed. The modification for the gluon and ghost propagators,
introduced by the vacuum state defined in the previous chapter,
are calculated.

\section{General Form of the Propagator}

As it is well known in the Quantum Field Theory to calculate any
element of the S-Matrix, after applying the reduction formulas, it
is necessary to obtain the vacuum expectation value of the
temporal ordering of Heisemberg operators \cite{Gasiorowicz}. That
is it is needed to calculate
\begin{equation}
\langle \Psi \mid T\left(\hat{A}_H\left(x_1\right) \hat{A}_H\left(
x_2\right) \hat{A}_H\left(x_3\right)...\right) \mid \Psi \rangle,
\label{orden1}
\end{equation}
where $\mid \Psi \rangle $ is the real vacuum of the interacting
theory. For simplifying the exposition it is considered a scalar
field, the generalization for vector fields is straightforward.

Using the relations between the operators in the Interaction and
Heisemberg representations
\begin{eqnarray*}
&&\hat{A}_H\left(x\right) =\hat{U}\left(0,t\right) \hat{A}_I\left(
x\right) \hat{U}\left(t,0\right), \\ &&\hat{U}\left(
t_1,t_2\right) \hat{U}\left(t_2,t_3\right) =\hat{U}\left(
t_1,t_3\right)
\end{eqnarray*}
and assuming that the real vacuum interacting state can be
obtained from the non-interacting one under the adiabatic
connection of the interaction. The expression (\ref{orden1}) takes
the form \cite{Gasiorowicz}
\begin{equation}
\frac{\langle \Phi \mid T\left\{ \hat{A}_I\left(x_1\right)
\hat{A}_I\left(x_2\right) \hat{A}_I\left(x_3\right)...\exp \left(
-\int\limits_{-\infty }^\infty H_i\left(t\right) dt\right)
\right\} \mid \Phi \rangle }{\langle \Phi \mid T\left\{ \exp
\left(-\int\limits_{-\infty }^\infty H_i\left(t\right) dt\right)
\right\} \mid \Phi \rangle }, \label{orden2}
\end{equation}
where $\Phi $ is the non interacting vacuum of the theory.

To evaluate these quantities it is needed to develop the
exponential in series of perturbation theory and calculate the
vacuum expectation values of the temporal ordering of fields in
the interaction representation ($ \hat{A}_I\left(x\right) $), but
in this representation the field operators are like free fields
($\hat{A}^0\left(x\right))$ about which much is known.

\[
\hat{A}_I\left(x\right) =\hat{A}^0\left(x\right).
\]
And it is necessary to evaluate terms of the form

\begin{equation}
\langle \Phi \mid T\left(\hat{A}^0\left(x_1\right) \hat{A}^0\left(
x_2\right) \hat{A}^0\left(x_3\right)...\right) \mid \Phi \rangle.
\end{equation}

Introducing the auxiliary generating functional

\begin{equation}
Z\left[ J\right] \equiv \langle \Phi \mid T\left(\exp \left\{
i\int d^4xJ\left(x\right) A^0\left(x\right) \right\} \right) \mid
\Phi \rangle, \label{genfun}
\end{equation}
it is possible to write for the relevant expectation values the
expression

\begin{equation}
\langle \Phi \mid T\left(\hat{A}^0\left(x_1\right) \hat{A}^0\left(
x_2\right) \hat{A}^0\left(x_3\right)...\right) \mid \Phi \rangle
=\left(\frac 1i\frac \delta {\delta J\left(x_1\right) }\frac
1i\frac \delta {\delta J\left(x_1\right) }\frac 1i\frac \delta
{\delta J\left(x_1\right) }...Z\left[ J\right] \right) _{J=0}.
\end{equation}

Considering now the auxiliary functional

\begin{equation}
Z\left[ J;t\right]\equiv \langle \Phi \mid T\left(\exp \left\{
i\int\limits_{-\infty }^tdt\int d^3xJ\left(x\right) A^0\left(
x\right) \right\} \right) \mid \Phi \rangle
\end{equation}
and defining $W\left(t\right)$ through the relation

\begin{eqnarray}
&&T\left(\exp \left\{ i\int\limits_{-\infty }^tdt\int d^3xJ\left(
x\right) A^0\left(x\right) \right\} \right)\nonumber \\ &&
=T\left(\exp \left\{ i\int\limits_{-\infty }^tdt\int d^3xJ\left(
x\right) A^{0-}\left(x\right) \right\} \right) W\left(t\right),
\label{T11}
\end{eqnarray}
where $A^{0-}\left(x\right)$ and \thinspace $A^{0+}\left(
x\right)$ are the negative (creation) and positive (annihilation)
frequency parts, respectively.

The $t$ differentiation on the expression (\ref{T11}), takes the
form
\begin{eqnarray}
&i&\int\limits_{x_0=t}d^3xJ\left(x\right) A^0\left(x\right)
T\left(\exp \left\{ i\int\limits_{-\infty }^tdt\int d^3xJ\left(
x\right) A^{0-}\left(x\right) \right\} \right) W\left(t\right)
\nonumber \\ &=&T\left(\exp \left\{ i\int\limits_{-\infty
}^tdt\int d^3xJ\left(x\right) A^{0-}\left(x\right) \right\}
\right) \frac{dW\left(t\right) }{dt}+ \nonumber \\
&&+i\int\limits_{x_0=t}d^3xJ\left(x\right) A^{0-}\left(x\right)
T\left(\exp \left\{ i\int\limits_{-\infty }^tdt\int d^3xJ\left(
x\right) A^{0-}\left(x\right) \right\} \right) W\left(t\right).
\end{eqnarray}

Keeping in mind that the free field creation operators commute,
for all times, the following relation holds

\begin{equation}
\left[ A^{0-}\left(x\right),A^{0-}\left(y\right) \right] =0,
\end{equation}
then the $T$ instruction can be eliminated and after some algebra
is obtained {\setlength\arraycolsep{0.5pt}
\begin{eqnarray}
\frac{dW\left(t\right) }{dt} &=&i\exp \left\{
-i\int\limits_{-\infty }^tdt\int d^3xJ\left(x\right) A^{0-}\left(
x\right) \right\} \int\limits_{x_0=t}d^3xJ\left(x\right)
A^{0+}\left(x\right) \times \nonumber \\ &&\times \exp \left\{
i\int\limits_{-\infty }^tdt\int d^3xJ\left(x\right) A^{0-}\left(
x\right) \right\} W\left(t\right) \nonumber \\
&=&i\int\limits_{y_0=t}d^3yJ\left(y\right) \left\{ A^{0+}\left(
y\right) -i\int\limits_{-\infty }^td^4xJ\left(x\right) \left[
A^{0-}\left(x\right),A^{0+}\left(y\right) \right] \right\} W\left(
t\right). \label{dif1}
\end{eqnarray}}

The initial condition on $W\left(t\right)$ is
\begin{equation}
W\left(-\infty \right) =1.
\end{equation}

Then the solution of (\ref{dif1}) is
\begin{eqnarray*}
W\left(t\right) &=&\exp \left\{ i\int\limits_{-\infty
}^td^4yJ\left(y\right) A^{0+}\left(y\right) \right\}\nonumber
\\&& \times \exp \left\{ \int\limits_{-\infty }^td^4y
\int\limits_{-\infty }^{y_0}d^4xJ\left(y\right) J\left(x\right)
\left[ A^{0-}\left(x\right),A^{0+}\left(y\right) \right] \right\},
\end{eqnarray*}
when $t\rightarrow \infty $ this expression takes the form
\begin{eqnarray*}
W\left(\infty \right) &=&\exp \left\{ i\int d^4yJ\left(y\right)
A^{0+}\left(y\right) \right\} \times \\ &&\times \exp \left\{ \int
d^4xd^4y\theta \left(y_0-x_0\right) J\left(y\right) J\left(
x\right) \left[ A^{0-}\left(x\right),A^{0+}\left(y\right) \right]
\right\}.
\end{eqnarray*}

Therefore, the generating functional (\ref{genfun}) can be written
in the following way \cite{Gasiorowicz}
\begin{eqnarray}
Z\left[ J\right] &\equiv &\langle \Phi \mid \exp \left\{ i\int
d^4xJ\left(x\right) A^{0-}\left(x\right) \right\} \exp \left\{
i\int d^4yJ\left(y\right) A^{0+}\left(y\right) \right\} \mid \Phi
\rangle \nonumber \\ &&\times \exp \left\{ \frac i2\int
d^4xd^4yJ\left(x\right) D(x-y)J\left(y\right) \right\},
\label{bosones}
\end{eqnarray}
where $D(x-y)$ is the usual propagator for an scalar particle.

In case that is needed to calculate a similar matrix element for
fermions the following functional is defined

\begin{equation}
Z\left[ \eta,\bar{\eta}\right] \equiv \langle \Phi \mid T\left(
\exp \left\{ i\int d^4x\left[ \bar{\eta}\left(x\right) \psi
^0\left(x\right) + \bar{\psi}^0\left(x\right) \eta \left(x\right)
\right] \right\} \right) \mid \Phi \rangle.
\end{equation}

Because of the anticommuting properties of $\bar{\psi},\ \psi$
fields the introduced sources $\bar{\eta},\ \eta$ satisfy
anticommuting relations between then and with the field operators.

\newpage

Here is assumed the left differentiation convention, then the
S-Matrix element can be calculate by the following expression

\begin{eqnarray}
&&\langle \Phi \mid T\left(\psi ^0\left(y_1\right)
\bar{\psi}^0\left(z_1\right) \psi ^0\left(y_2\right)
...\bar{\psi}^0\left(z_k\right) \right) \mid \Phi \rangle
\nonumber \\ &&=\left(\frac 1i\frac \delta {\delta \eta \left(
z_k\right) }...\frac 1i\frac \delta {\delta \bar{\eta}\left(
y_2\right) }\frac 1i\frac \delta {\delta \eta \left(z_1\right)
}\frac 1i\frac \delta {\delta \bar{\eta} \left(y_1\right) }Z\left[
\eta,\bar{\eta}\right] \right) _{\eta,\bar{\eta} =0}.
\end{eqnarray}
Now, in the same way that for the bosons, the following auxiliary
functional is defined by

\begin{equation}
Z\left[ \eta,\bar{\eta};t\right] \equiv \langle \Phi \mid T\left(
\exp \left\{ i\int\limits_{-\infty }^tdt\int d^3x\left[
\bar{\eta}\left(x\right) \psi ^0\left(x\right) +\bar{\psi}^0\left(
x\right) \eta \left(x\right) \right] \right\} \right) \mid \Phi
\rangle
\end{equation}
and the corresponding $G\left(t\right)$ functional by

\begin{eqnarray}
&&T\left(\exp \left\{ i\int\limits_{-\infty }^tdt\int d^3x\left[
\bar{\eta} \left(x\right) \psi ^0\left(x\right)
+\bar{\psi}^0\left(x\right) \eta \left(x\right) \right] \right\}
\right) \nonumber \\ &&=T\left(\exp \left\{ i\int\limits_{-\infty
}^tdt\int d^3x\left[ \bar{\eta} \left(x\right) \psi ^{0-}\left(
x\right) +\bar{\psi}^{0-}\left(x\right) \eta \left(x\right)
\right] \right\} \right) G\left(t\right). \label{ferm1}
\end{eqnarray}

Manipulations completely parallel to those leading to (\ref{dif1})
give {\setlength\arraycolsep{0.5pt}
\begin{eqnarray}
\frac{dG\left(t\right) }{dt}=i &&\left[
\int\limits_{y_0=t}d^3y\left(\bar{ \eta}\left(y\right) \psi
^{0+}\left(y\right) +\bar{\psi}^{0+}\left(y\right) \eta \left(
y\right) \right) \right. \nonumber \\ &&+i\int\limits_{-\infty
}^td^4x\int\limits_{y_0=t}d^3y\bar{\eta}\left(y\right) \left\{
\psi ^{0+}\left(y\right),\bar{\psi}^{0-}\left(x\right) \right\}
\eta \left(x\right) \nonumber \\ &&\left. -i\int\limits_{-\infty
}^td^4x\int\limits_{y_0=t}d^3y\bar{\eta} \left(x\right) \left\{
\psi ^{0-}\left(x\right),\bar{\psi}^{0+}\left(y\right) \right\}
\eta \left(y\right) \right] G\left(t\right),
\end{eqnarray}}
This equation is easily integrated to obtain the solution
\begin{eqnarray}
G\left(t\right) &=&\exp \left\{ i\int\limits_{-\infty
}^td^4y\left(\bar{ \eta}\left(y\right) \psi ^{0+}\left(y\right)
+\bar{\psi}^{0+}\left(y\right) \eta \left(y\right) \right)
\right\} \nonumber \\ &&\times \exp \left\{ -\int\limits_{-\infty
}^td^4y\int\limits_{-\infty }^{y_0}d^4x\bar{\eta}\left(y\right)
\left\{ \psi ^{0+}\left(y\right),\bar{ \psi}^{0-}\left(x\right)
\right\} \eta \left(x\right) \right\} \nonumber
\\ &&\times \exp \left\{ \int\limits_{-\infty
}^td^4x\int\limits_{-\infty }^{x_0}d^4y\bar{\eta}\left(y\right)
\left\{ \psi ^{0-}\left(y\right),\bar{ \psi}^{0+}\left(x\right)
\right\} \eta \left(x\right) \right\},
\end{eqnarray}
where in the last term the dummy variables $x$ and $y$ were
interchanged. Consequently the following expression for the
generating functional arise \cite{Gasiorowicz}
{\setlength\arraycolsep{0.5pt}
\begin{eqnarray}
Z\left[ \eta,\bar{\eta}\right] &\equiv &\langle \Phi \mid \exp
\left\{ i\int d^4x\left[ \bar{\eta}\left(x\right) \psi ^{0-}\left(
x\right) +\bar{ \psi}^{0-}\left(x\right) \eta \left(x\right)
\right] \right\} \nonumber \\ &&\quad\times\exp \left\{ i\int
d^4x\left[ \bar{\eta}\left(x\right) \psi ^{0+}\left(x\right)
+\bar{\psi}^{0+}\left(x\right) \eta \left(x\right) \right]
\right\} \mid \Phi \rangle \nonumber \\ &&\times \exp \left\{
i\int d^4xd^4y\bar{\eta}\left(x\right) S\left(x-y\right) \eta
\left(y\right) \right\} \label{fermiones}
\end{eqnarray}}
where $S\left(x-y\right)$ is the standard fermion propagator.

As much for the case of bosons as for fermions the term related
with the vacuum expectation value for the usual vacuum is one.
This is so because the annihilation operators are located to the
right and to the left those of creation. However in the present
work the vacuum expectation values generate the propagator
modifications, because the vacuum state considered is not the
trivial one. The other term in the generating functional
expression, that is expressed by a simple exponential of c
numbers, gives the usual propagator and it has the same form when
is calculated by this operational method or alternatively by the
functional method.

Then, starting from the analysis in the present section it can be
concluded that from an operation formalism point of view any
modification to the usual propagators is only generated by a
change in the vacuum state of the theory. And these modifications
can be determined through the vacuum expectation values in
(\ref{bosones}) and (\ref{fermiones}). From a functional formalism
point of view, the propagator modifications are generated by
changes in the boundary conditions.

\section{Modified Gluon Propagator}

As it follows from the general form of the Wick Theorem, analyzed
in the previous section, the modification of the gluon propagator
introduced by the modified vacuum state (\ref{Vacuum}) is defined
by the expression

\begin{equation}
\langle \widetilde{\Phi }\mid \exp \left\{ i\int d^4xJ^{\mu
,a}\left(x\right) A_\mu ^{a-}\left(x\right) \right\} \exp \left\{
i\int d^4xJ^{\mu,a}\left(x\right) A_\mu ^{a+}\left(x\right)
\right\} \mid \widetilde{\Phi } \rangle, \label{mod}
\end{equation}
for each value of the color index $a$. All the different colors
can be worked out independently because of the commutation
relations between the annihilation and creation operators for the
free theory. At the necessary point of the analysis all the color
contributions will be included.

The annihilation and creation fields in (\ref{mod}) are given by

\begin{eqnarray*}
A_\mu ^{a+}\left(x\right) &=&\sum\limits_{\vec{k}}\left(
\sum\limits_{\sigma =1,2}A_{\vec{k},\sigma }^af_{k,\mu }^\sigma
\left(x\right) +A_{\vec{k}}^{L,a}f_{k,L,\mu }\left(x\right)
+B_{\vec{k} }^af_{k,S,\mu }\left(x\right) \right), \\ A_\mu
^{a-}\left(x\right) &=&\sum\limits_{\vec{k}}\left(
\sum\limits_{\sigma =1,2}A_{\vec{k},\sigma }^{a+}f_{k,\mu
}^{\sigma *}\left(x\right) +A_{\vec{k}}^{L,a+}f_{k,L,\mu
}^{*}\left(x\right) +B_{\vec{k} }^{a+}f_{k,S,\mu }^{*}\left(
x\right) \right).
\end{eqnarray*}

In what follows it is calculated explicitly, for each color, the
action of the exponential operators
\begin{eqnarray}
&&\exp \left\{ i\int d^4xJ^{\mu,a}\left(x\right) A_\mu ^{a+}\left(
x\right) \right\} \mid \Phi \rangle \nonumber \\ &&=\exp \left\{
i\int d^4xJ^{\mu,a}\left(x\right) \sum\limits_{\vec{k} }\left(
\sum\limits_{\sigma =1,2}A_{\vec{k},\sigma }^af_{k,\mu }^\sigma
\left(x\right) +A_{\vec{k}}^{L,a}f_{k,L,\mu }\left(x\right)
+B_{\vec{k} }^af_{k,S,\mu }\left(x\right) \right) \right\}
\nonumber \\ &&\quad\times \exp \left\{ \sum\limits_{\sigma
=1,2}\frac 12C_\sigma \left(\left| \vec{p}\right| \right)
A_{\vec{p},\sigma }^{a+}A_{\vec{p},\sigma }^{a+}+C_3\left(\left|
\vec{p}\right| \right) \left(B_{\vec{p}}^{a+}A_{
\vec{p}}^{L,a+}+i\overline{c}_{\vec{p}}^{a+}c_{\vec{p}}^{a+}\right)
\right\} \mid 0\rangle. \label{expd}
\end{eqnarray}

After a systematic use of the commutation relations among the
annihilation and creation operators, the exponential operators can
be decomposed in products of exponential for each space-time mode.
This fact allows to perform the calculation for each mode
independently. Then the expression (\ref {expd}) takes the form
{\setlength\arraycolsep{0.2pt}
\begin{eqnarray}
&\prod\limits_{\sigma =1,2}&\exp \left\{ i\int d^4xJ^{\mu
,a}\left(x\right) \sum\limits_{\vec{k}}A_{\vec{k},\sigma
}^af_{k,\mu }^\sigma \left(x\right) \right\} \exp \left\{ \frac
12C_\sigma \left(\left| \vec{p}\right| \right) A_{\vec{p},\sigma
}^{a+}A_{\vec{p},\sigma }^{a+}\right\} \mid 0\rangle \nonumber \\
&\times &\exp \left\{ i\int d^4xJ^{\mu,a}\left(x\right)
\sum\limits_{\vec{k }}\left(B_{\vec{k}}^af_{k,S,\mu }\left(
x\right)+A_{\vec{k} }^{L,a}f_{k,L,\mu }\left(x\right)\right)
\right\} \nonumber \\ &\times &\exp \left\{ C_3\left(\left|
\vec{p}\right| \right) B_{\vec{p} }^{a+}A_{\vec{p}}^{L,a+}\right\}
\mid 0\rangle \exp \left\{ C_3\left(\left| \vec{p}\right| \right)
i\overline{c}_{\vec{p}}^{a+}c_{\vec{p}}^{a+}\right\} \mid
0\rangle.
\end{eqnarray}}

For a transverse component, it is necessary to calculate

\begin{equation}
\exp \left\{ i\int d^4xJ^{\mu,a}\left(x\right)
\sum\limits_{\vec{k}}A_{ \vec{k},\sigma }^af_{k,\mu }^\sigma
\left(x\right) \right\} \exp \left\{ \frac 12C_\sigma \left(
\left| \vec{p}\right| \right) A_{\vec{p},\sigma
}^{a+}A_{\vec{p},\sigma }^{a+}\right\} \mid 0\rangle \
\text{\qquad for } \sigma =1,2 \label{modT}
\end{equation}

The following recourse is used to calculate this expression;
calling $U$ the first exponential in (\ref{modT}) this expression
can be written as

\begin{equation}
\exp \left\{ \frac 12C_\sigma\left(p\right) \left(
UA_{\vec{p},\sigma }^{a+}U^{-1}\right) \left(UA_{\vec{p},\sigma
}^{a+}U^{-1}\right) \right\} \mid 0\rangle, \label{modTt}
\end{equation}
since
\[
U^{-1}\mid 0\rangle =\mid 0\rangle.
\]

The inverse $U^{-1}$ is the same $U$ when in the exponential
argument the sign is changed. Using the Baker-Hausdorf formula

\[
\exp [\hat{F}]\hat{G}\exp [-\hat{F}]=\exp \left\{ [\hat{F},\
]\right\} \hat{G }=\sum \frac 1{n!}\left[ \hat{F},\left[
\hat{F},....,\left[ \hat{F},\hat{G} \right].....\right] \right]
\]
and noticing that only the first and the second term in the
expansion are non-vanishing when $\hat{F}$ and $\hat{G}$ are
linear functions of annihilation and creation operators, it
follows

\[
\exp [\hat{F}]\hat{G}\exp [-\hat{F}]=\hat{G}+\left[
\hat{F},\hat{G}\right].
\]

Therefore, for the relevant commutators appearing in (\ref{modTt})
it follows
\[
\left[ i\int d^4xJ^{\mu,a}\left(x\right)
\sum\limits_{\vec{k}}A_{\vec{k},\sigma }^af_{k,\mu }^\sigma \left(
x\right),A_{\vec{p},\sigma }^{a+}\right] =i\int d^4xJ^{\mu
,a}\left(x\right) f_{p,\mu }^\sigma \left(x\right).
\]
Then for the expression (\ref{modT}) the following result is
obtained
\begin{equation}
\exp \left\{ \frac 12C_\sigma \left(\left| \vec{p}\right| \right)
\left(A_{ \vec{p},\sigma }^{a+}+i\int d^4xJ^{\mu,a}\left(x\right)
f_{p,\mu }^\sigma \left(x\right) \right) ^2\right\} \mid 0\rangle
\ \label{T}
\end{equation}

For the longitudinal and scalar modes, following the above
procedure, the result obtained is
\begin{eqnarray}
&&\exp \left\{i\int d^4xJ^{\mu,a}\left(x\right)
\sum\limits_{\vec{k} }\left(B_{\vec{k}}^af_{k,S,\mu }\left(
x\right) +A_{\vec{k} }^{L,a}f_{k,L,\mu }\left(x\right) \right)
\right\}\nonumber \\ && \times \exp \left\{ C_3\left(\left|
\vec{p}\right| \right) B_{\vec{p}}^{a+}A_{\vec{p}}^{L,a+}\right\}
\mid 0\rangle \nonumber
\\ &&=\exp \left\{ C_3\left(\left| \vec{p}\right| \right) \left(
B_{\vec{p} }^{a+}-i\int d^4xJ^{\mu,a}\left(x\right) f_{p,L,\mu
}\left(x\right) \right) \right. \nonumber
\\&& \qquad\qquad\qquad\qquad \times \left.
\left(A_{\vec{p}}^{L,a+} -i\int d^4xJ^{\mu,a}\left(x\right)
f_{p,S,\mu }\left(x\right) \right) \right\} \mid
0\rangle.\label{L}
\end{eqnarray}
where the expressions below were used
\begin{eqnarray*}
&&\left[ \left(i\int d^4xJ^{\mu,a}\left(x\right)
\sum\limits_{\vec{k} } A_{\vec{k} }^{L,a}f_{k,L,\mu }\left(
x\right) \right),B_{\vec{p}}^{a+}\right] =-i\int
d^4xJ^{\mu,a}\left(x\right) f_{p,L,\mu }\left(x\right) \\ &&\left[
\left(i\int d^4xJ^{\mu,a}\left(x\right) \sum\limits_{\vec{k} }
B_{\vec{k}}^af_{k,S,\mu }\left(x\right) \right),A_{\vec{p}
}^{L,a+}\right] =-i\int d^4xJ^{\mu,a}\left(x\right) f_{p,S,\mu
}\left(x\right).
\end{eqnarray*}

For the full modification calculation (\ref{mod}), it is necessary
to evaluate
\begin{equation}
\langle \Phi \mid \exp \left\{ i\int d^4xJ^{\mu,a}\left(x\right)
A_\mu ^{a-}\left(x\right) \right\} =\left(\exp \left\{ -i\int
d^4xJ^{\mu,a}\left(x\right) A_\mu ^{a+}\left(x\right) \right\}
\mid \Phi \rangle \right) ^{\dagger }, \label{left}
\end{equation}
which can be easily obtained by conjugating the result for the
right hand side, through (\ref{T}) and (\ref{L}).

Then, substituting (\ref{T}), (\ref{L}) and (\ref{left}) in
(\ref{mod}), the following expression should be calculated

\begin{eqnarray}
&&\frac 1N\langle 0\mid \exp \left\{ \sum\limits_{\sigma
=1,2}\frac 12C_\sigma ^{*}\left(\left| \vec{p}\right| \right)
\left(A_{\vec{p},\sigma }^a+i\int d^4xJ^{\mu,a}\left(x\right)
f_{p,\mu }^{\sigma *}\left(x\right) \right) ^2\right\} \nonumber
\\ &&\qquad \times \exp \left\{ \sum\limits_{\sigma =1,2}\frac
12C_\sigma \left(\left| \vec{p}\right| \right) \left(
A_{\vec{p},\sigma }^{a+}+i\int d^4xJ^{\mu,a}\left(x\right)
f_{p,\mu }^\sigma \left(x\right) \right) ^2\right\} \mid 0\rangle
\nonumber \\ &&\times\langle 0\mid \exp \left\{ C_3^{*}\left(
\left| \vec{p}\right| \right) \left(B_{\vec{p}}^a-i\int d^4xJ^{\mu
,a}\left(x\right) f_{p,L,\mu }^{*}\left(x\right) \right) \right.
\nonumber \\ &&\qquad\qquad\qquad\qquad\qquad \times \left. \left(
A_{\vec{p}}^{L,a}-i\int d^4xJ^{\mu,a}\left(x\right) f_{p,S,\mu
}^{*}\left(x\right) \right) \right\} \nonumber \\ &&\qquad \times
\exp \left\{ C_3\left(\left| \vec{p}\right| \right) \left(B_{
\vec{p}}^{a+}-i\int d^4xJ^{\mu,a}\left(x\right) f_{p,L,\mu
}\left(x\right) \right) \right. \nonumber \\
&&\qquad\qquad\qquad\qquad\qquad \times \left. \left(
A_{\vec{p}}^{L,a+}-i\int d^4xJ^{\mu,a}\left(x\right) f_{p,S,\mu
}\left(x\right) \right) \right\} \mid 0\rangle \nonumber \\
&&\times\langle 0\mid \exp \left(-iC_3^{*}\left(\left|
\vec{p}\right| \right)
c_{\vec{p}}^a\overline{c}_{\vec{p}}^a\right) \exp \left(
iC_3\left(\left| \vec{p}\right| \right)
\overline{c}_{\vec{p}}^{a+}c_{\vec{p}}^{a+}\right) \mid 0\rangle
\label{mod1}
\end{eqnarray}

In the expression (\ref{mod1}) the calculated contribution, for
each transverse mode, is
\begin{equation}
\exp \left\{ -\int \frac{d^4xd^4y}{2Vp_0}J^{\mu,a}\left(x\right)
J^{\nu,a}\left(y\right) \epsilon _\mu ^\sigma \left(p\right)
\epsilon _\nu ^\sigma \left(p\right) \frac{\left(C_\sigma \left(
\left| \vec{p}\right| \right) +C_\sigma ^{*}\left(\left|
\vec{p}\right| \right) +2\left| C_\sigma \left(\left|
\vec{p}\right| \right) \right| ^2\right) }{2\left(1-\left|
C_\sigma \left(\left| \vec{p}\right| \right) \right| ^2\right)
}\right\}, \label{T1}
\end{equation}
and the longitudinal and scalar mode contribution is

\begin{equation}
\exp \left\{ -\int \frac{d^4xd^4y}{2Vp_0}J^{\mu,a}\left(x\right)
J^{\nu,a}\left(y\right) \epsilon _{S,\mu }\left(p\right) \epsilon
_{L,\nu }\left(p\right) \frac{\left(C_3\left(\left| \vec{p}\right|
\right) +C_3^{*}\left(\left| \vec{p}\right| \right) +2\left|
C_3\left(\left| \vec{p} \right| \right) \right| ^2\right) }{\left(
1-\left| C_3\left(\left| \vec{p} \right| \right) \right| ^2\right)
}\right\}, \label{L1}
\end{equation}

The detailed analysis of these calculations can be found in the
Appendixes 1 and 2.

Therefore, after collecting the contributions of all the modes,
assuming $ C_1\left(\left| \vec{p}\right| \right) =C_2\left(
\left| \vec{p}\right| \right) =C_3\left(\left| \vec{p}\right|
\right) $ (which follows necessarily in order to obtain Lorentz
invariance) and using the properties of the defined vectors basis,
the modification to the propagator becomes

\begin{equation}
\exp \left\{ \frac 12\int \frac{d^4xd^4y}{2p_0V}J^{\mu,a}\left(
x\right) J^{\nu,a}\left(y\right) g_{\mu \nu }\left[ \frac{\left(
C_1\left(\left| \vec{p}\right| \right) +C_1^{*}\left(\left|
\vec{p}\right| \right) +2\left| C_1\left(\left| \vec{p}\right|
\right) \right| ^2\right) }{\left(1-\left| C_1\left(\left|
\vec{p}\right| \right) \right| ^2\right) }\right] \right\}.
\label{totmodF}
\end{equation}

In the expression (\ref{totmodF}), the combination of the
$C_1\left(\left| \vec{p}\right| \right) $ constant is always real
and nonnegative, for all $ \left| C_1\left(\left| \vec{p}\right|
\right) \right| <1$.

Now it is possible to perform the limit process
$\vec{p}\rightarrow 0$. In doing this limit, it is considered that
each component of the linear momentum $\vec{p}$ is related with
the quantization volume by

\[
p_x\sim \frac 1a,\ p_y\sim \frac 1b,\ p_z\sim \frac 1c,\ V=abc\sim
\frac 1{\left| \vec{p}\right| ^3},
\]
$\ \ $

And it is necessary to calculate

\begin{equation}
\lim_{\vec{p}\rightarrow 0}\frac{\left(C_1\left(\left|
\vec{p}\right| \right) +C_1^{*}\left(\left| \vec{p}\right| \right)
+2\left| C_1\left(\left| \vec{p}\right| \right) \right| ^2\right)
}{4p_0V\left(1-\left| C_1\left(\left| \vec{p}\right| \right)
\right| ^2\right) },\label{Lim1}
\end{equation}

Then, after fixing a dependence of the arbitrary constant $C_1$ of
the form $ \left| C_1\left(\left| \vec{p}\right| \right) \right|
\sim \left(1-\kappa \left| \vec{p}\right| ^2\right),\kappa >0$,
and $C_1\left(0\right) \neq -1$ the limit (\ref{Lim1}) becomes

\begin{equation}
\lim_{\vec{p}\rightarrow 0}\frac{\left(C_1\left(\left|
\vec{p}\right| \right) +C_1^{*}\left(\left| \vec{p}\right| \right)
+2\left| C_1\left(\left| \vec{p}\right| \right) \right| ^2\right)
\left| \vec{p}\right| ^3\frac 1{\left(1-\left(1-\kappa \left|
\vec{p}\right| ^2\right) ^2\right) }}{4p_0}=\frac C{2\left(2\pi
\right) ^4}
\end{equation}
where $C$ is an arbitrary real and nonnegative constant,
determined by the also real and nonnegative constant $\kappa$.

Therefore, the total modification to the propagator including all
color values turns to be {\setlength\arraycolsep{0.5pt}
\begin{eqnarray}
&\prod\limits_{a=1,..,8}&\langle \widetilde{\Phi }\mid \exp
\left\{ i\int d^4xJ^{\mu,a}\left(x\right) A_\mu ^{a-}\left(
x\right) \right\} \exp \left\{ i\int d^4xJ^{\mu,a}\left(x\right)
A_\mu ^{a+}\left(x\right) \right\} \mid \widetilde{\Phi }\rangle
\nonumber \\ &=&\exp \left\{ \sum\limits_{a=1,..8}\int
d^4xd^4yJ^{\mu,a}\left(x\right) J^{\nu,a}\left(y\right) g_{\mu \nu
}\frac C{2\left(2\pi \right) ^4}\right\}.
\end{eqnarray}}

The generating functional associated to the proposed initial
state, including the usual perturbative piece for $\alpha =1$, can
be written in the form

\begin{equation}
Z[J]=\exp \left\{ \frac i2\sum\limits_{a,b=1,..8}\int
d^4xd^4yJ^{\mu,a}\left(x\right) \widetilde{D}_{\mu \nu
}^{ab}(x-y)J^{\nu,b}\left(y\right) \right\},
\end{equation}
where

\begin{equation}
\widetilde{D}_{\mu \nu }^{ab}(x-y)=\int \frac{d^4k}{\left(2\pi
\right) ^4} \delta ^{ab}g_{\mu \nu }\left[ \frac 1{k^2}-iC\delta
\left(k\right) \right] \exp \left\{ -ik\left(x-y\right) \right\}
\label{propag}
\end{equation}
which shows that the gluon propagator has the same form proposed
in \cite{Cabo}, for the selected gauge parameter value $\alpha =1$
(which corresponds to $\alpha =-1$ in that reference).

\section{Modified Ghost Propagator}

In the present section the possible modification to the ghost
propagator will be analyzed. As was shown in Sec. 3.1 for
fermionic particles the expression for the modification,
introduced by a nontrivial vacuum state, is

\begin{eqnarray}
&&\langle \widetilde{\Phi } \mid \exp \left\{ i\int d^4x\left(
\overline{\xi }^a\left(x\right) c^{a-}\left(x\right)
+\overline{c}^{a-}\left(x\right) \xi ^a\left(x\right) \right)
\right\} \nonumber \\&&\qquad \times \exp \left\{ i\int d^4x\left(
\overline{\xi }^a\left(x\right) c^{a+}\left(x\right)
+\overline{c}^{a+}\left(x\right) \xi ^a\left(x\right) \right)
\right\} \mid \widetilde{\Phi }\rangle, \label{ini}
\end{eqnarray}
where

\begin{eqnarray}
c^{a+}\left(x\right)
&=&\sum\limits_{\vec{k}}c_{\vec{k}}^ag_k\left(x\right),\qquad
c^{a-}\left(x\right) =\sum\limits_{\vec{k}}c_{\vec{k}
}^{a+}g_k^{*}\left(x\right), \nonumber \\ \overline{c}^{a+}\left(
x\right) &=&\sum\limits_{\vec{k}}\overline{c}_{\vec{k
}}^ag_k\left(x\right),\qquad \overline{c}^{a-}\left(x\right)
=\sum\limits_{\vec{k}}\overline{c}_{\vec{k}}^{a+}g_k^{*}\left(
x\right).
\end{eqnarray}

Now it is calculated explicitly the action of the exponential
operator
\begin{eqnarray}
&&\exp \left\{ i\int d^4x\left(\overline{\xi }^a\left(x\right)
c^{a+}\left(x\right) +\overline{c}^{a+}\left(x\right) \xi ^a\left(
x\right) \right) \right\} \exp \left\{ C_3\left(\left|
\vec{p}\right| \right) i\overline{c}_{\vec{p}}^{a+}
c_{\vec{p}}^{a+}\right\} \mid 0\rangle \nonumber \\ &&=\left(
1+i\int d^4y\overline{\xi }^a\left(y\right) \sum\limits_{\vec{k}
^{\prime }}c_{\vec{k}^{\prime }}^ag_{k^{\prime }}\left(y\right)
\right) \left(1+i\int d^4x\sum\limits_{\vec{k}}
\overline{c}_{\vec{k}}^ag_k\left(x\right) \xi ^a\left(x\right)
\right) \times \nonumber \\ &&\qquad \times \left(1+C_3\left(
\left| \vec{p}\right| \right) i\overline{c}_{
\vec{p}}^{a+}c_{\vec{p}}^{a+}\right) \mid 0\rangle. \label{Ghost}
\end{eqnarray}

The grassman character of the field and sources allowed expanding
the exponential retaining only the first two terms in the
expansion.

With the use of the following relations

{\
\begin{eqnarray}
\overline{\xi }^a\left(y\right) c_{\vec{k}^{\prime
}}^a\overline{c}_{\vec{p} }^{a+}c_{\vec{p}}^{a+}\mid 0\rangle
&=&i\delta _{\vec{k}^{\prime },\vec{p}} \overline{\xi }^a\left(
y\right) c_{\vec{p}}^{a+}\mid 0\rangle, \nonumber
\\ \overline{c}_{\vec{k}}^a\xi ^a\left(x\right)
\overline{c}_{\vec{p}}^{a+}c_{ \vec{p}}^{a+}\mid 0\rangle
&=&i\delta _{\vec{k},\vec{p}}\overline{c}_{\vec{p} }^{a+}\xi
^a\left(x\right) \mid 0\rangle, \nonumber \\ \overline{\xi
}^a\left(y\right) c_{\vec{k}^{\prime }}^ai\delta _{\vec{k},
\vec{p}}\overline{c}_{\vec{p}}^{a+}\xi ^a\left(x\right) \mid
0\rangle &=&-\delta _{\vec{k},\vec{p}}\delta _{\vec{k}^{\prime
},\vec{p}}\overline{ \xi }^a\left(y\right) \xi ^a\left(x\right)
\mid 0\rangle,
\end{eqnarray}
}the expression (\ref{Ghost}) can be written as
\begin{eqnarray}
&&\left[ 1+C_3\left(\left| \vec{p}\right| \right) \left(
i\overline{c}_{ \vec{p}}^{a+}c_{\vec{p}}^{a+}-i\int d^4xg_p\left(
x\right) \left(\overline{ \xi }^a\left(x\right)
c_{\vec{p}}^{a+}+\overline{c}_{\vec{p}}^{a+}\xi ^a\left(x\right)
\right) \right. \right. + \nonumber \\ &&\left. \left. +i\int
d^4y\int d^4xg_p\left(y\right) g_p\left(x\right) \overline{\xi
}^a\left(y\right) \xi ^a\left(x\right) \right) \right] \mid
0\rangle. \label{rhs}
\end{eqnarray}

In addition the formula
\begin{eqnarray}
&&\langle \widetilde{\Phi }\mid \exp \left\{ i\int d^4x\left(
\overline{\xi } ^a\left(x\right) c^{a-}\left(x\right)
+\overline{c}^{a-}\left(x\right) \xi ^a\left(x\right) \right)
\right\} \nonumber \\ &&=\left[ \exp \left\{ i\int d^4x\left(
\overline{\xi }^{a\dagger }\left(x\right) c^{a+}\left(x\right)
+\overline{c}^{a+}\left(x\right) \xi ^{a\dagger }\left(x\right)
\right) \right\} \mid \widetilde{\Phi }\rangle \right] ^{\dagger
}, \label{rsh1}
\end{eqnarray}
allows to calculate the left hand side of (\ref{ini}) using
(\ref{rhs}).

Then the expression (\ref{ini}), substituting (\ref{rhs}) and
(\ref{rsh1}), takes the form {\setlength\arraycolsep{0.5pt}
\begin{eqnarray}
&\langle 0\mid &\left[ 1-C_3^{*}\left(\left| \vec{p}\right|
\right) \left(ic_{\vec{p}}^a\overline{c}_{\vec{p}}^a-i\int
d^4xg_p^{*}\left(x\right) \left(c_{\vec{p}}^a\overline{\xi
}^a\left(x\right) +\xi ^a\left(x\right)
\overline{c}_{\vec{p}}^a\right) \right. \right. \nonumber \\
&&\left. \left. +i\int d^4y\int d^4xg_p^{*}\left(y\right)
g_p^{*}\left(x\right) \xi ^a\left(y\right) \bar{\xi}^a\left(
x\right) \right) \right] \nonumber \\ &\times &\left[ 1+C_3\left(
\left| \vec{p}\right| \right) \left(i\overline{c
}_{\vec{p}}^{a+}c_{\vec{p}}^{a+}-i\int d^4xg_p\left(x\right)
\left(\overline{\xi }^a\left(x\right)
c_{\vec{p}}^{a+}+\overline{c}_{\vec{p} }^{a+}\xi ^a\left(x\right)
\right) \right. \right. \nonumber \\ &&\left. \left. +i\int
d^4y\int d^4xg_p\left(y\right) g_p\left(x\right) \overline{\xi
}^a\left(y\right) \xi ^a\left(x\right) \right) \right] \mid
0\rangle. \label{ghomod}
\end{eqnarray}}

In this case, the expression (\ref{ghomod}) calculus is easier
than the one realized for gluons. And the result of its
contribution, canceling out the normalization factor, is

\begin{equation}
\exp \left[ \frac{i\int d^4xd^4y\overline{\xi }^a\left(x\right)
\xi ^a\left(y\right) \left(C_3\left(\left| \vec{p}\right| \right)
+C_3^{*}\left(\left| \vec{p}\right| \right) -2\left| C_3\left(
\left| \vec{p }\right| \right) \right| ^2\right) }{2Vp_0\left(
1-\left| C_3\left(\left| \vec{p}\right| \right) \right| ^2\right)
}\right],
\end{equation}
which in the limit $\vec{p}\rightarrow 0$, under the same
condition considered for the gluon modification limit, takes the
form

\begin{equation}
\exp \left\{ -\sum\limits_{a=1,..8}i\int d^4xd^4y\overline{\xi
}^a\left(x\right) \xi ^a\left(y\right) \frac{C_G}{\left(2\pi
\right) ^4}\right\}.
\end{equation}
In this expression $C_G$ is a real and nonnegative constant. It is
interesting to note that choosing $C_3\left(0\right) =1$, then
$C_G=0$ and there is no modification to the ghost propagator as
was chosen in the previous work \cite{Cabo}.

The ghost generating functional associated to the proposed initial
state, including the usual perturbative piece for $\alpha =1$, can
be written in the form

\begin{equation}
Z_G[\overline{\xi },\xi ]=\exp \left\{
i\sum\limits_{a,b=1,..8}\int d^4xd^4y \overline{\xi }^a\left(
x\right) \widetilde{D}_G^{ab}(x-y)\xi ^b\left(y\right) \right\},
\end{equation}
where
\begin{equation}
\widetilde{D}_G^{ab}(x-y)=\int \frac{d^4k}{\left(2\pi \right)
^4}\delta ^{ab}\left[ \frac{\left(-i\right) }{k^2}-C_G\delta
\left(k\right) \right] \exp \left\{ -ik\left(x-y\right) \right\}.
\end{equation}

\chapter{Summary}

By using the operational formulation for Quantum Gauge Fields
Theory developed by Kugo and Ojima, a particular state vector for
QCD in the non-interacting limit, that obeys the BRST physical
state condition, was constructed. The general motivation for
looking this wave function is to search for a reasonably good
description of low energy QCD properties, through giving
foundation to the perturbative expansion proposed in \cite{Cabo}.
The high energy QCD description should not be affected by the
modified perturbative initial state. In addition it can be
expected that the adiabatic connection of the color interaction
starting with it as an initial condition, generate at the end the
true QCD interacting ground state. In case of having the above
properties, the analysis would allow to understand the real vacuum
as a superposition of infinite number of soft gluon pairs.

It has been checked that properly fixing the free parameters in
the constructed state, the perturbation expansion proposed in the
former work \cite{Cabo} is reproduced for the special value
$\alpha =1$ of the gauge constant. Therefore, the appropriate
gauge is determined for which the expansion introduced in that
work is produced by an initial state, satisfying the physical
state condition for the BRST quantization procedure. The fact that
the non-interacting initial state is a physical one, lead to
expect that the final wave-function after the adiabatic connection
of the color interaction will also satisfy the physical state
condition for the interacting theory. If this assumption is
correct, the results for calculations of transition amplitudes and
the values of physical quantities should be also physically
meaningful. In future, a quantization procedure for arbitrary
values of $\alpha$ will be also considered. It is expected that
with its help the gauge parameter independence of the physical
quantities could be implemented. It seems possible that the
results of this generalization will lead to $\alpha $ dependent
polarizations for gluons and ghosts and their respective
propagators, which however could produce $\alpha $ independent
results for the physical quantities. However, this discussion will
be delayed for future consideration.

It is important to mention now a result obtained during the
calculation of the gluon propagator modification, in the chosen
construction. It is that the arbitrary constant $C$ is determined
here to be real and nonnegative. This outcome restricts an
existing arbitrariness in the discussion given in the previous
work. As this quantity $C$ is also determining the square of the
generated gluon mass as positive or negative, real or imaginary,
therefore it seems very congruent to arrive to a definite
prediction of $C$ as real and positive.

The modification to the standard free ghost propagator introduced
by the proposed initial state, was also calculated. Moreover,
after considering the free parameter in the proposed trial state
as real, which it seems the most natural assumption, the ghost
propagator is not be modified, as it was assumed in \cite{Cabo}.

Some tasks which can be addressed in future works are: The study
of the applicability of the Gell-Mann and Low theorem with respect
to the adiabatic connection of the interaction, starting from the
here proposed initial state. The investigation of zero modes
quantization, that is gluon states with exact vanishing four
momentum. The ability to consider them with success would allow a
formally cleaner definition of the proposed state, by excluding
the auxiliary momentum $\vec{p}$ recursively used in the
construction carry out. Finally, the application of the proposed
perturbation theory in the study of some problems related with
confinement and the hadron structure.

\appendix

\chapter{Transverse Mode Contribution}

The transverse mode contribution is determined by the expression

\begin{eqnarray}
&&\langle 0\mid \exp \left\{ \frac 12C_\sigma ^{*}\left(\left|
\vec{p} \right| \right) \left(A_{\vec{p},\sigma }^a+i\int
d^4xJ^{\mu,a}\left(x\right) f_{p,\mu }^{\sigma *}\left(x\right)
\right) ^2\right\} \nonumber
\\ &&\quad\times \exp \left\{ \frac 12C_\sigma \left(\left|
\vec{p}\right| \right) \left(A_{\vec{p},\sigma }^{a+}+i\int
d^4xJ^{\mu,a}\left(x\right) f_{p,\mu }^\sigma \left(x\right)
\right) ^2\right\} \mid 0\rangle \label{A1}
\end{eqnarray}

For simplifying the exposition, the following notation is
introduced

\begin{eqnarray}
C^{*} &\equiv &C_\sigma ^{*}\left(\left| \vec{p}\right| \right),\
C\equiv C_\sigma \left(\left| \vec{p}\right| \right),\text{ \qquad
}\hat{A} ^{+}\equiv A_{\vec{p},\sigma }^{a+},\ \hat{A}\equiv
A_{\vec{p},\sigma }^a \text{\ }, \nonumber \\ a_1 &\equiv &i\int
d^4xJ^{\mu,a}\left(x\right) f_{p,\mu }^{\sigma
*}\left(x\right),\text{ \qquad }a_2\equiv i\int d^4xJ^{\mu
,a}\left(x\right) f_{p,\mu }^\sigma \left(x\right). \label{nota1}
\end{eqnarray}

Then the expression (\ref{A1}) takes the form

\begin{eqnarray}
&&\langle 0 \mid \exp \left\{ \frac 12C^{*}\left(
\hat{A}+a_1\right) ^2\right\} \exp \left\{ \frac 12C\left(
\hat{A}^{+}+a_2\right) ^2\right\} \mid 0\rangle \label{A2}
\\ &&=\exp \left\{ \frac{C^{*}}2a_1^2+\frac C2a_2^2\right\}
\langle 0\mid \exp \left\{ \frac{C^{*}}2\hat{A}^2
+C^{*}a_1\hat{A}\right\} \exp \left\{ \frac C2
\hat{A}^{+2}+Ca_2\hat{A}^{+}\right\} \mid 0\rangle. \nonumber
\end{eqnarray}

For the action of the exponential, linear in the annihilation
operator, at the left on the right the result obtained is
\begin{eqnarray}
&&\exp \left\{ C^{*}a_1\hat{A}\right\} \exp \left\{ \frac
C2\hat{A} ^{+2}+Ca_2\hat{A}^{+}\right\} \mid 0\rangle \nonumber
\\ &&=\exp \left\{ \frac C2\left(\hat{A}^{+}+C^{*}a_1\right)
^2+Ca_2\left(\hat{A}^{+}+C^{*}a_1\right) \right\} \mid 0\rangle,
\label{A3}
\end{eqnarray}
where the same procedure used for calculating (\ref{modT}) is
considered.

The expression (\ref{A2}), considering (\ref{A3}), can be written
in the form

\begin{equation}
\exp \left\{ \frac{C^{*}}2a_1^2+\frac C2\left(
C^{*}a_{1}+a_2\right) ^2\right\} \langle 0\mid \exp \left\{
\frac{C^{*}}2\hat{A}^2\right\} \exp \left\{ \frac
C2\hat{A}^{+2}+C\hat{A}^{+}\left(C^{*}a_{1}+a_2\right) \right\}
\mid 0\rangle \label{A4}
\end{equation}

It is possible in (\ref{A4}) to act with the exponential linear in
the creation operator at the right on the left and the result is

\begin{eqnarray}
&&\exp \left\{ \frac{C^{*}}2a_1^2+\frac C2\left(
C^{*}a_{1}+a_2\right) ^2\left(1+\left| C\right| ^2\right) \right\}
\nonumber \\ &&\times \langle 0 \mid \exp \left\{
\frac{C^{*}}2\hat{A}^2+C^{*}C\left(C^{*}a_{1}+a_2\right)
\hat{A}\right\} \exp \left\{ \frac C2\hat{A} ^{+2}\right\} \mid
0\rangle \label{A5}
\end{eqnarray}

In such a way after n-steps it is possible to arrive to a
recurrence relation, which can be proven by mathematical
induction. This recurrence relation has the form

\begin{eqnarray}
&&\exp \left\{ \frac{C^{*}}2a_1^2+\frac C2\left(
C^{*}a_{1}+a_2\right) ^2\sum\limits_{m=0}^n\left[ \left| C\right|
^{2\left(2m\right) }+\left| C\right| ^{2\left(2m+1\right) }\right]
\right\} \nonumber \\ &&\times \langle 0 \mid \exp \left\{
\frac{C^{*}}2\hat{A}^2+C^{*n+1}C^{n+1} \left(
C^{*}a_{1}+a_2\right) \hat{A}\right\} \exp \left\{ \frac C2\hat{A}
^{+2}\right\} \mid 0\rangle \label{rec1}
\end{eqnarray}

Lets probe it, acting with the exponential linear in the
annihilation operator at the left on the right the result is

\begin{eqnarray}
&&\exp \left\{ \frac{C^{*}}2a_1^2+\frac C2\left(
C^{*}a_{1}+a_2\right) ^2\left(\sum\limits_{m=0}^n\left[ \left|
C\right| ^{2\left(2m\right) }+\left| C\right| ^{2\left(
2m+1\right) }\right] +\left| C\right| ^{4\left(n+1\right) }\right)
\right\} \nonumber \\ &&\times \langle 0 \mid \exp \left\{
\frac{C^{*}}2\hat{A}^2\right\} \exp \left\{ \frac
C2\hat{A}^{+2}+C^{*n+1}C^{n+2}\left(C^{*}a_{1}+a_2\right)
\hat{A}^{+}\right\} \mid 0\rangle,
\end{eqnarray}
now acting on the left with the exponential linear in the creation
operator is obtained the relation

\begin{eqnarray}
&&\exp \left\{ \frac{C^{*}}2a_1^2+\frac C2\left(
C^{*}a_{1}+a_2\right) ^2\sum\limits_{m=0}^{n+1}\left[ \left|
C\right| ^{2\left(2m\right) }+\left| C\right| ^{2\left(
2m+1\right) }\right] \right\} \nonumber \\ &&\times \langle 0 \mid
\exp \left\{ \frac{C^{*}}2\hat{A}^2+C^{*n+2}C^{n+2} \left(
C^{*}a_{1}+a_2\right) \hat{A}\right\} \exp \left\{ \frac C2\hat{A}
^{+2}\right\} \mid 0\rangle \label{A6}
\end{eqnarray}
which probe the recurrence relation (\ref{rec1}).

At this point the limit $n\rightarrow \infty$ is taken,
considering $\left| C\right| <1$ which implies that

\begin{eqnarray}
&&\lim_{n\rightarrow \infty }\left| C\right| ^{2n}=0, \nonumber
\\ &&\lim_{n\rightarrow \infty }\sum\limits_{m=0}^n\left[ \left|
C\right| ^{2\left(2m\right) }+\left| C\right| ^{2\left(
2m+1\right) }\right] =\frac 1{\left(1-\left| C\right| ^2\right) },
\label{lim}
\end{eqnarray}
and the expression (\ref{A6}) in this limit has the form,

\begin{equation}
\exp \left\{ \frac{\left(C^{*}a_{1}^2+Ca_2^2+2C^{*}Ca_1a_2\right)
}{2\left(1-\left| C\right| ^2\right) }\right\} \langle 0\mid \exp
\left\{ \frac{C^{*}} 2\hat{A}^2\right\} \exp \left\{ \frac
C2\hat{A}^{+2}\right\} \mid 0\rangle \label{A7}
\end{equation}

Finally, the notation (\ref{nota1}) is substituted in (\ref{A7}).
After that, the functions of $\vec{p}$ are expanded in the
vicinity of $\vec{p}=0$, keeping in mind that the sources are
located in a space finite region it is necessary to consider only
the first terms in the expansion. Then for the expression
(\ref{A7}) it is obtained the result (\ref{T1}), the
renormalization factors cancel out.

\chapter{Longitudinal and Scalar Modes Contribution}

The longitudinal and scalar modes contribution is determined by
the expression

\begin{eqnarray}
&&\langle 0\mid \exp \left\{ C_3^{*}\left(\left| \vec{p}\right|
\right) \left(B_{\vec{p}}^a-i\int d^4xJ^{\mu,a}\left(x\right)
f_{p,L,\mu }^{*}\left(x\right) \right) \right. \nonumber \\
&&\qquad\qquad\qquad\qquad\qquad \times \left. \left(
A_{\vec{p}}^{L,a}-i\int d^4xJ^{\mu,a}\left(x\right) f_{p,S,\mu
}^{*}\left(x\right) \right) \right\} \nonumber \\ &&\qquad \times
\exp \left\{ C_3\left(\left| \vec{p}\right| \right) \left(B_{
\vec{p}}^{a+}-i\int d^4xJ^{\mu,a}\left(x\right) f_{p,L,\mu
}\left(x\right) \right) \right. \nonumber \\
&&\qquad\qquad\qquad\qquad\qquad \times \left. \left(
A_{\vec{p}}^{L,a+}-i\int d^4xJ^{\mu,a}\left(x\right) f_{p,S,\mu
}\left(x\right) \right) \right\} \mid 0\rangle \label{B1}
\end{eqnarray}
introducing the following notation,

\begin{eqnarray}
C^{*} &\equiv &C_3^{*}\left(\left| \vec{p}\right| \right),\
C\equiv C_3\left(\left| \vec{p}\right| \right),\text{ \ \
}\hat{A}^{+}\equiv A_{ \vec{p}}^{L,a+},\ \hat{A}\equiv
A_{\vec{p}}^{L,a},\text{ \ \ }\hat{B} ^{+}\equiv
B_{\vec{p}}^{a+},\ \hat{B}\equiv B_{\vec{p}}^a, \nonumber \\ a_1
&\equiv &-i\int d^4xJ^{\mu,a}\left(x\right) f_{p,S,\mu }^{*}\left(
x\right),\text{ \qquad }a_2\equiv -i\int d^4xJ^{\mu,a}\left(
x\right) f_{p,S,\mu }\left(x\right), \nonumber \\ b_1 &\equiv
&-i\int d^4xJ^{\mu,a}\left(x\right) f_{p,L,\mu }^{*}\left(
x\right),\text{ \qquad }b_2\equiv -i\int d^4xJ^{\mu,a}\left(
x\right) f_{p,L,\mu }\left(x\right), \label{nota2}
\end{eqnarray}
the expression (\ref{B1}) takes the form
\begin{eqnarray}
&&\langle 0\mid \exp \left\{ C^{*}\left(\hat{A}+a_1\right) \left(
\hat{B} +b_1\right) \right\} \exp \left\{ C\left(
\hat{B}^{+}+b_2\right) \left(\hat{A}^{+}+a_2\right) \right\} \mid
0\rangle \nonumber \\ &&=\exp \left\{ C^{*}a_1b_1+Ca_2b_2\right\}
\langle 0\mid \exp \left\{ C^{*}\left(\hat{A}\hat{B}
+b_1\hat{A}+a_1\hat{B}\right) \right\} \nonumber
\\ &&\text{ \qquad \qquad \qquad \qquad \qquad \qquad }\times \exp
\left\{ C\left(\hat{B}^{+}\hat{A}^{+}+
b_2\hat{A}^{+}+a_2\hat{B}^{+}\right) \right\} \mid 0\rangle
\label{B2}
\end{eqnarray}

For the action of the exponential linear in the annihilation
operator at the left on the right, is obtained

\begin{eqnarray}
&&\exp \left\{ C^{*}\left(b_1\hat{A}+a_1\hat{B}\right) \right\}
\exp \left\{ C\left(
\hat{B}^{+}\hat{A}^{+}+b_2\hat{A}^{+}+a_2\hat{B}^{+}\right)
\right\} \mid 0\rangle \label{B3} \\ &&=\exp \left\{ C\left[
\left(\hat{B}^{+}-C^{*}b_1\right) \left(\hat{A}
^{+}-C^{*}a_1\right) +b_2\left(\hat{A}^{+}-C^{*}a_1\right)
+a_2\left(\hat{B }^{+}-C^{*}b_1\right) \right] \right\} \mid
0\rangle, \nonumber
\end{eqnarray}
the same procedure used for calculating (\ref{L}) is considered.

Following the same steps described in the previous appended for
transverse modes, in this case the recurrence relation obtained
for longitudinal and scalar modes is
{\setlength\arraycolsep{0.5pt}
\begin{eqnarray}
&&\exp \left\{ C^{*}a_1b_1+C\left(C^{*}a_1-a_2\right) \left(
C^{*}b_1-b_2\right) \sum\limits_{m=0}^n\left[ \left| C\right|
^{2\left(2m\right) }+\left| C\right| ^{2\left(2m+1\right) }\right]
\right\}\label{B4} \\ &&\langle 0\mid \exp \left\{
C^{*}\hat{A}\hat{B}+C^{*n+1}C^{n+1}\left(\left(
C^{*}b_1-b_2\right) \hat{A}+\left(C^{*}a_1-a_2\right)
\hat{B}\right) \right\} \exp \left\{
C\hat{B}^{+}\hat{A}^{+}\right\} \mid 0\rangle.\nonumber
\end{eqnarray}}

For the expression (\ref{B4}), in the limit $n\rightarrow \infty $
considering $\left| C\right| <1$, the following relation is
obtained

\begin{eqnarray}
&&\exp \left\{ C^{*}a_1b_1+C\left(C^{*}a_1-a_2\right) \left(
C^{*}b_1-b_2\right) \frac 1{\left(1-\left| C\right| ^2\right)
}\right\} \nonumber \\ &&\quad \langle 0\mid \exp \left\{
C^{*}\hat{A}\hat{B}\right\} \exp \left\{ C
\hat{B}^{+}\hat{A}^{+}\right\} \mid 0\rangle. \label{B5}
\end{eqnarray}

Finally, the notation (\ref{nota2}) is substituted in (\ref{B5}),
the functions of $\vec{p}$ are expanded in the vicinity of
$\vec{p}=0$, and the result (\ref{L1}) is obtained.

\end{document}